# CAHIERS
# FRANÇOIS VIETE

Série I – N°8

2004

*«Nouvelle théorie des taches du soleil »*
*Esprit Pezenas (1692-1776), S.J.*
*Archives départementales de l'Hérault, Ms. D.12S, S.D.*
*(C. 1766-70), FOL. 261-267.*

Édition commentée et annotée
par Guy Boistel

# "Nouvelle théorie des taches du Soleil", Esprit Pézenas

## Archives départementales de l'Hérault, Ms. D.128, S.D. (C. 1766-70), FOL. 261-267

*"New theory of sunspots", Esprit Pézenas*

**Guy Boistel (dir.)**

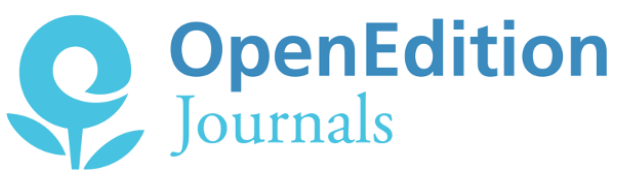

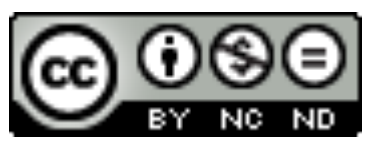

# INTRODUCTION DE LA PUBLICATION

L'astronome et hydrographe marseillais Esprit Pézenas rédige le manuscrit publié ici entre 1768 et 1772. Ses observations régulières des taches solaires lui permettent de préciser, entre autres, la période de rotation du Soleil. L'appareil critique proposé par Guy Boistel comporte une biographie de l'auteur, le contexte scientifique et politique du manuscrit composé après l'expulsion des jésuites, et des pièces annexes, dont les rapports à l'Académie des sciences.

The Marseille astronomer and hydrographer Esprit Pézenas wrote the manuscript published here between 1768 and 1772. His regular observations of sunspots enabled him to specify, among other things, the period of rotation of the Sun. The critical apparatus proposed by Guy Boistel includes a biography of the author, the scientific and political context of the manuscript composed after the expulsion of the Jesuits, and appendices, including reports to the Academy of Sciences.

# CAHIERS
# FRANÇOIS VIETE

Série I – N°8

2004

*«Nouvelle théorie des taches du soleil »*
*Esprit Pezenas (1692-1776), S.J.*
*Archives départementales de l'Hérault, Ms. D.12S, S.D.*
*(C. 1766-70), FOL. 261-267.*

Édition commentée et annotée
par Guy Boistel

# COMMENTAIRE

Guy BOISTEL

## Préambule

Le manuscrit « Nouvelle théorie des taches du Soleil » a été identifié et attribué à l'astronome et hydrographe jésuite marseillais Esprit Pezenas (1692-1776) au cours de l'année 2002, lors de la recherche systématique des papiers le concernant dans les différents fonds français.

Pour des raisons techniques, l'édition de ce manuscrit n'est imprimée qu'en 2012 alors qu'elle paraît dans un numéro des *Cahiers François Viète* millésimé 2004. Cette édition fait suite à la publication en 2003 de l'inventaire des manuscrits et des œuvres imprimées du père Pezenas dans la *Revue d'histoire des sciences*[1].

## Les débuts de la physique solaire au XVII^e^ siècle et le problème scientifique de la détermination de la rotation du Soleil par l'observation de ses taches

C'est entre les années 1610 et 1620 que, grâce aux développements successifs de la lunette astronomique, plusieurs astronomes ont contribué à la mise en évidence de la rotation du Soleil sur lui-même à l'aide des observations de ses taches. L'Anglais Thomas Harriot (1560-1621) a observé les premières taches solaires en décembre 1610. Johannes Fabricius (1587-1616) en Hollande, fut le premier à comprendre que le mouvement des taches solaires d'un jour à l'autre était dû à la rotation du

[1] Guy Boistel, 2003, « Inventaire chronologique des œuvres imprimées et manuscrites du père Esprit Pezenas (1692-1776), jésuite, astronome et hydrographe marseillais », *Revue d'histoire des sciences*, vol. 56/1, 221-245.

Soleil sur lui-même. Enfin, il revient au père jésuite Christoph Scheiner (1575-1650) et à Galilée (1564-1642) d'avoir observé et étudié systématiquement l'apparition et le mouvement des taches. Avec une petite lunette astronomique, il est aisé d'observer le déplacement et la rotation apparente des taches solaires (qui se traduit par un décalage d'environ 13° par jour vers l'Ouest en coordonnées héliographiques). En supposant que ces taches appartiennent à la surface solaire – ce qui constitue alors un véritable débat en soi[2] – et compte tenu des imprécisions des observations et d'une théorie du mouvement du Soleil encore incomplète[3], les astronomes du début du XVII$^{e}$ siècle ont trouvé une période de rotation du Soleil sur lui-même comprise entre 26 et 29 jours et demi (soit un mois lunaire). Il faut attendre le début du XIX$^{e}$ siècle pour que l'astronome allemand Heinrich Schwabe (1789-1875) mette en évidence un cycle d'activité solaire de onze ans, caractérisé par une variation cyclique du nombre de taches solaires présentes sur la surface du Soleil[4].

Le problème scientifique posé par le mouvement des taches solaires n'est pas si simple. Les taches apparaissent vers les latitudes élevées du disque solaire (35 à 45°), de manière symétrique dans les deux hémisphères, et glissent progressivement vers l'équateur solaire. Mais le Soleil n'est pas une sphère solide. Il présente une rotation différentielle : des couches de latitudes différentes ne tournent pas à la même vitesse, les taches tournant plus vite à l'équateur. Sur un diagramme montrant l'évolution de la latitude des taches en fonction du temps, on voit apparaître des formes, les fameuses « ailes de papillon ». Par ailleurs, les premières déterminations de la durée de rotation du Soleil ne tenaient pas compte de l'inclinaison de l'équateur du Soleil sur l'écliptique. C'est au père Scheiner que l'on doit la découverte de cette l'inclinaison (égale à environ 7°)[5]. En 1764, Lalande montre que les mesures ne sont toujours pas suffisamment précises et appelle de tous ses vœux de meilleures déterminations de cette inclinaison par les astronomes[6].

---

[2] Dans son essai intitulé *Il Saggiatore* (1619), Galilée revendiqua la paternité de la découverte des taches solaires sur le P. Scheiner ; s'ensuivirent des discussions se déplaçant assez vite sur les terrains philosophique et théologique. Sur Harriot, Galilée et Scheiner, lire la passionnante saga de Walter M. Mitchell, 1916, « The history of the discovery of the solar spots », *Popular Astronomy*, 1916, 22-30 ; 82-96 ; 149-162 ; 206-218 ; 290-303 : 341-354 ; 428-441 ; 488-499 ; 562-570 (disponibles sur l'abstract service de la Nasa ADS). Voir aussi : Bernard Dame, 1966, « Galilée et les taches solaires (1610-1613) », *Revue d'histoire des sciences et de leurs applications*, 19/4, 307-370 ; William R. Shea, 1970, « Galileo, Scheiner, and the interpretation of Sunspots », *Isis*, 61, 498-519.

[3] La mécanique céleste pré-newtonienne est encore suffisamment imprécise pour apporter de grandes incertitudes dans ce genre de calculs.

[4] Judit Brody, 2002, *The enigma of sunspots. A story of discovery and scientific revolution*, Edinburgh, Floris Books, complément récent des références données en note 3. Sur Schwabe : Franz Flury, 1927, « L'astronome amateur Schwabe », *Bulletin de l'Observatoire de Lyon*, 1927, 86-91.

Le mouvement progressif des taches vers l'équateur solaire est relativement lent et, pendant l'observation du déplacement d'une tache d'un bord à l'autre du Soleil, la Terre s'est aussi déplacée autour de ce dernier. L'observateur n'observe donc pas les taches sous le même angle vu de la Terre entre deux observations rapprochées de quelques jours. Ainsi, entre le début des observations des taches solaires dans les années 1610 et le milieu du XVIII$^{e}$ siècle, quelques astronomes ont proposé des méthodes, souvent graphiques (le P. Scheiner, Johannes Hevelius (1611-1687), Jean-Dominique Cassini (1625-1712), Joseph-Nicolas Delisle (1688-1768), Jacques Cassini (1677-1756), le P. Esprit Pezenas (1692-1776), Jérôme Lalande (1732-1807) notamment) puis des méthodes plus analytiques (Guillaume de Saint-Jacques de Silvabelle (1722-1801), Lalande ou Pierre-Achille Dionis Duséjour (1734-1794) en particulier) afin de résoudre ces différentes questions en tenant compte ou non, à des degrés divers, du déplacement de la Terre autour du Soleil pendant les observations. Jérôme Lalande donne un très bon aperçu historique de ces différentes tentatives dans le tome 2 de la première édition de son *Astronomie* (Paris, 1764)[7].

Notons que le traitement des observations des taches solaires par les différents astronomes conduit indirectement au calcul de la période de rotation ; il consiste tout d'abord, et surtout, à déterminer l'inclinaison de l'équateur solaire sur l'écliptique, dont le complémentaire est l'inclinaison de l'axe de rotation du Soleil sur ce même écliptique. Enfin, la période de rotation est souvent accessoirement déduite des divers calculs et pourrait presque passer pour anecdotique. Tous les astronomes s'entendent sur une durée d'environ 27 jours à quelques heures près, en plus ou en moins.

---

[5] Christopher Scheiner, 1630, *Rosa Ursina*, numérisé et mis en ligne sur le site du Musée des sciences de Florence :

http://fermi.imss.fi.it/rd/bdv?/bdviewer/bid=367767.

André Danjon, 1994, *Astronomie générale. Astronomie sphérique et éléments de mécanique céleste*, Paris, A. Blanchard, 349-351 : l'inclinaison de l'axe solaire est environ de 7°,25. La rotation sidérale (conventionnelle) du Soleil est de 25,38 jours ; la rotation synodique correspondante est de 27,275 jours.

[6] Jérôme Lalande, 1764, *Astronomie*, Paris, tome II, p. 1220.

[7] J. Lalande, 1764, *op. cit.*, tome II, art. 2502 et suiv., 1204 -1222.

L'esprit géométrique des savants des Lumières est plus souvent préoccupé par l'aspect théorique d'une méthode, son style mathématique, que par ses applications numériques. Seuls les véritables astronomes, observateurs au fait des nouveautés en matière de mathématiques, s'appliquent à fournir des données numériques et les résultats de leurs méthodes.

En outre, avec le développement de l'optique instrumentale au XVIII<sup>e</sup> siècle qui voit la construction de miroirs de bronze poli et de lunettes aux verres d'une qualité optique de plus en plus grande, les astronomes disposent d'instruments leur permettant de discerner les taches dans l'atmosphère de la planète Jupiter. Ainsi, les méthodes développées pour le traitement des taches solaires servent-elles aussi à la détermination de la période de rotation de Jupiter et de toute autre planète présentant des taches. Lalande montre bien comment ces méthodes peuvent aussi servir au traitement de la libration de la Lune par l'observation des « taches » lunaires, telles que ses cratères et/ou ses mers[8].

Enfin, l'observation et le dessin des taches bénéficient aussi de l'amélioration des techniques d'observation à la fin du XVII<sup>e</sup> siècle avec le perfectionnement du micromètre par l'astronome Adrien Auzout (1622-1691) à l'Observatoire royal de Paris, et son emploi systématique par l'abbé Jean Picard (1620-1682). Celui-ci incorpore à l'oculaire de la lunette astronomique un micromètre à fil mobile de Auzout. Picard et son élève Philippe de la Hire (1667-1719) observent systématiquement au quart de cercle le diamètre angulaire apparent du Soleil, atteignant une précision proche de la seconde d'arc. La Hire et ses successeurs emploieront les déterminations des durées de passages des bords du Soleil et de ses taches au méridien, grâce aux gains sensibles des horloges astronomiques, qui tiennent désormais la seconde[9].

Toutes ces observations et ces techniques nouvelles contribuent, encore maintenant, à une meilleure connaissance des éléments de base de la physique solaire et de sa variabilité. Il a été ainsi possible de mettre en évidence une anomalie dans l'évolution du nombre de taches entre 1645 et 1705, appelée « minimum de Maunder »[10], se traduisant notamment par une

---

[8] J. Lalande, 1764, *op. cit.*, 1222 et suiv.

[9] Guy Picolet (dir.), 1987, *Jean Picard et les débuts de l'astronomie de précision au XVII<sup>e</sup> siècle*, Paris, CNRS.

[10] En hommage à l'astronome anglais Edward Maunder (1851-1928) qui a étendu les études historiques entreprises par l'allemand Gustav Spörer (1822-1895). Ce dernier est à l'origine du « minimum de Spörer » qui semble être survenu entre 1420 et 1570, mis en évidence grâce aux variations d'abondance de l'isotope 14 du carbone dans les anneaux de croissance des arbres corrélées à

dissymétrie marquée et réelle de répartition des taches sur les deux hémisphères solaires. Des études plus fines sur les époques encadrant ce minimum de Maunder pour lesquelles nous disposons d'observations de taches solaires[11], révèlent deux phénomènes. Premièrement, ce minimum semble s'accompagner d'une baisse de l'activité solaire et de la température moyenne terrestre et paraît corroborer l'existence d'un « petit âge glaciaire » entre 1550 et 1850 déduite de la variation d'abondance de l'isotope 14 du carbone. En second lieu, les observations du diamètre angulaire du Soleil semblent montrer une variation de la rotation angulaire du Soleil, celle-ci s'accélérant sensiblement lors du minimum de Maunder. Ces deux observations sont également contestées dans le cadre des débats sur la climatologie, et nous n'irons pas plus loin sur ces questions[12].

On imagine alors aisément l'importance de la publication de collections d'observations et de dessins des taches solaires, – comme par exemple, l'*Histoire céleste ou recueil de toutes les observations astronomiques faites par ordre du Roy*, publiée en 1741 par l'astronome Pierre-Charles Le Monnier, dont il sera question plus loin –, ou de toute étude quantitative sur la rotation du Soleil.

C'est dans ce cadre scientifique que nous présentons le manuscrit, en grande partie inédit, de la « Nouvelle théorie des taches du Soleil » écrit par l'astronome et professeur d'hydrographie jésuite marseillais, le père Esprit Pezenas (1692-1776), texte composé et revu entre les années 1766 et 1772. Dans ce manuscrit, dont nous allons étudier les conditions de sa composition, le P. Pezenas donne l'une des dernières méthodes géométriques, apparentées aux méthodes graphiques, permettant de déduire l'inclinaison de l'équateur du Soleil sur l'écliptique, à l'aide de trois

---

l'activité solaire. L'isotope C-14 est produit par réaction des neutrons du vent solaire et de l'azote-14 ; mais plus l'activité solaire est intense et moins il y a de C-14 produit dans la haute atmosphère, car le vent solaire dévie les rayons cosmiques qui produisent le C-14.

[11] Scheiner débute ses observations 20 ans avant le minimum de Maunder, Hevelius juste au début et celui-ci totalise près de 4000 jours d'observations continues. Picard et La Hire observent durant le minimum ; La Hire et ses élèves lors de la reprise de l'activité solaire en 1710. On dispose ensuite de nombreuses observations, plus ou moins régulières, des jésuites notamment, publiées dans les *Mémoires de Trévoux*, au cours du XVIII$^{e}$ siècle.

[12] D.V. Hoyt & K.H. Schatten, 1997, *The role of the Sun in climate change,* Oxford University Press. Voir aussi Andrew E. Dessler & Edward A. Parson, 2006, *The Science and Politics of global climate change. A guide to the Debate*, Cambridge University Press.

observations d'une tache solaire, à des époques données, et en tenant compte du mouvement de la Terre par rapport au Soleil pendant la durée des observations. Ce texte comporte tous les éléments numériques permettant de suivre et de comprendre l'application des méthodes de traitement géométrique des observations et des techniques d'observations citées plus haut dans l'introduction : passages des bords et des taches du Soleil au méridien, emploi des micromètres objectifs et des micromètres à fils. Au détour de ses calculs, le P. Pezenas annonce une durée d'environ 26 jours et 9 heures pour la rotation du Soleil autour de son axe, commente et rectifie quelques données d'observations publiées dans l'*Histoire céleste* de Le Monnier.

L'étude du P. Pezenas est donc éclairante à plus d'un titre. Elle illustre parfaitement la pratique d'un astronome des Lumières soucieux d'être compris de ses lecteurs et montre comment un astronome, connaissant parfaitement la littérature de sa discipline, tire le meilleur parti des recueils d'observations astronomiques.

## Introduction à l'édition de la « Nouvelle théorie des taches du Soleil » du P. Pezenas

Originaire d'une famille noble d'Avignon, Esprit Pezenas suit la formation traditionnelle des jésuites au Collège de cette ville et développe de solides compétences en mathématiques. Il devient l'un des plus efficaces prédicateurs jésuites auprès des populations provençales. Esprit Pezenas se voit confier en 1728 la direction de l'observatoire des jésuites de Marseille de la maison Sainte-Croix, située à la montée des Accoules, sur la rive nord du vieux port. L'observatoire était inoccupé depuis le décès de son directeur le P. Thioly en 1720, provoqué par la terrible peste qui sévit en Provence en 1721-22, causant des milliers de morts. Pezenas devient professeur d'hydrographie en 1728 auprès des officiers des Galères royales et le restera jusqu'en 1749 (date de la suppression des galères).

Après un voyage de neuf mois à Paris cette même année, au cours duquel il noue de précieuses relations avec des membres de l'Académie royale des sciences, ainsi qu'avec le célèbre libraire-éditeur Antoine Jombert et quelques membres influents de la Cour, Esprit Pezenas devient le directeur du nouvel « observatoire royal de la Marine » à Marseille. À presque soixante ans, il est élu correspondant de l'astronome Joseph-Nicolas Delisle pour l'Académie des sciences, puis correspondant de la toute nouvelle Académie de Marine créée à Brest par le ministre Louis-Antoine Rouillé, et nommé recteur de la maison des jésuites de Sainte-Croix. Conservant sa pension royale de professeur d'hydrographie, Pezenas obtient des crédits destinés à la rénovation de l'observatoire.

Au cours des années 1750, il parvient à l'équiper de nouveaux télescopes à miroirs de bronze de James Short, et de micromètres objectifs de Dollond notamment, instruments coûteux et représentant alors ce qui se fait de mieux en matière d'instrumentation astronomique. Pezenas et son équipe (les jésuites Louis Lagrange et Jean-Baptiste Blanchard entre autres) développent un programme de recherche orienté très clairement vers une meilleure maîtrise de l'optique instrumentale (astronomique et nautique), la recherche de comètes, l'observation des taches solaires, la théorie des tables de la Lune et la traduction d'ouvrages de mathématiques de langue anglaise[13]. À la fin des années 1750, attirés par la réputation du P. Pezenas, des jésuites polonais et espagnols séjournent à l'observatoire de Marseille pour se former, soit aux observations astronomiques, soit à la traduction d'ouvrages de mathématiques. L'observatoire préfigure ainsi un centre de formation jésuite de haut niveau scientifique[14].

Malheureusement, la dispersion de la Compagnie de Jésus, qui a lieu en Provence durant l'Hiver 1762-63[15], met un terme brutal à ce développement scientifique. Pezenas rejoint définitivement Avignon en 1766. Avec les appuis dont il bénéficie au plus haut niveau, et contrairement à ses coreligionnaires, il peut poursuivre son activité scientifique et éditoriale, ainsi que son apostolat, jusqu'à son décès en 1776. Notons la publication d'une *Astronomie des marins* (1766), de la traduction et l'adaptation du *Cours complet d'optique* de Robert Smith en deux volumes assortie de nombreuses additions originales de Pezenas (1767), de la traduction de *La montre [de marine] de John Harrison* (1767)[16], et de quelques textes importants sur la détermination des longitudes en mer par les méthodes lunaires[17].

[13] G. Boistel, 2005, « L'observatoire des jésuites de Marseille sous la direction du P. Pezenas (1728-1763) », in G. Boistel (dir.), *Observatoires et patrimoine astronomique français*, in Cahiers d'histoire et de philosophie des sciences, n°54 , SFHST/ENS-LSH, ENS Éditions, Lyon, 27-45.

[14] G. Boistel, 2010, « Esprit Pezenas (1692-1776), jésuite, astronome et traducteur : un acteur méconnu de la diffusion de la science anglaise en France au XVIII^e^ siècle », in B. Joly & R. Fox (éds.), *Échanges entre savants français et britanniques depuis le XVII^e^ siècle*, Cahiers de logique et d'épistémologie n°7, Oxford, College Publications, 135-157.

[15] Rappelons qu'en 1761 le Parlement de Paris prend le prétexte de la banqueroute financière du Père jésuite Lavalette à La Martinique pour rendre la Compagnie de Jésus toute entière solidaire de ses créances et attaquer sévèrement leurs Constitutions. Les parlements des Provinces suivront avec un peu de retard les décisions du Parlement de Paris qui conduiront à la dispersion de l'Ordre puis à son interdiction temporaire en 1773 (la Compagnie renaît peu à peu après 1814).

## La « Nouvelle théorie des taches solaires » du P. Esprit Pezenas

Examinons le contexte de la composition et de la publication partielle de ce manuscrit non daté.

L'observation des taches solaires est, depuis l'affaire Galilée, une activité traditionnelle chez les astronomes jésuites : les cahiers d'observatoires connus comportent de très nombreuses observations et dessins de taches solaires[18]. Au cours des années 1750, les jésuites marseillais et leurs élèves, dont le jeune Guillaume de Saint-Jacques de Silvabelle, développent ce type d'observation à l'aide des nouveaux instruments qui équipent progressivement l'observatoire à partir de l'année 1752 : un grand télescope de type grégorien permettant un grossissement de 800 fois, ainsi que deux plus petits télescopes (Cassegrain et grégorien), avec leurs héliomètres-objectifs de Dollond[19].

Le 19 mars 1763, Pezenas est dépossédé de la direction de l'observatoire suite à une véritable descente de police à laquelle prend part Silvabelle, élève réputé prodige après que celui-ci a remarqué quelques erreurs dans le *Traité sur la précession des équinoxes* de d'Alembert[20]. Issu de la noblesse provençale, Saint-Jacques de Silvabelle est un élève régulier de l'observatoire dès le début des années 1750. Pezenas intervient personnellement et fait jouer ses amitiés académiques pour promouvoir autant qu'il lui est possible le *Traité des variations célestes* de Silvabelle (dont il nous reste une copie de la main même de Pezenas)[21]. Silvabelle observe comètes et taches du Soleil aux côtés des jésuites de Sainte-Croix au cours des années 1750 jusqu'en 1762, et les quelques notes d'observations et correspondances qui nous restent, nous donnent l'impression de relations amicales et intellectuelles sincères entre lui et les jésuites de Sainte-Croix. À la suite de la dispersion des jésuites, Silvabelle obtient le brevet de directeur de l'observatoire royal de Marseille en 1764, assorti d'une pension de 1200 livres ; il occupe cette charge jusqu'en 1801, bien que l'observatoire passe sous la tutelle de l'Académie des sciences lettres et arts de Marseille en 1781[22].

---

[16] Le prix britannique pour la mise au point d'une méthode de détermination des longitudes en mer est remis en 1765 et partagé entre l'horloger John Harrison, constructeur de la célèbre montre marine H4, et l'astronome allemand Tobias Mayer pour de nouvelles tables de la Lune, autorisant l'emploi de la méthode des distances lunaires en mer. Voir G. Boistel, 2001, *L'astronomie nautique au XVIII^e^ siècle en France : tables de la Lune et longitudes en mer*, thèse de doctorat, Université de Nantes (3 vols.) ; éditée en 2003 par l'A.N.R.T., Lille-3, 2 vols.

[17] G. Boistel, 2003 et 2010, *op. cit.* Voir aussi G. Boistel, 2001, « Deux documents inédits des PP. jésuites R.J. Boscovich et Esprit Pezenas sur les longitudes en mer », *Revue d'histoire des sciences*, 54/3, 383-397 ; ainsi que G. Boistel, 2002, « Les longitudes en mer au XVIII^e^ siècle sous le regard critique du père Pezenas », in Vincent Jullien (Dir.), *Le calcul des longitudes. Un enjeu pour les mathématiques, l'astronomie, la mesure du temps et la navigation*, Rennes, Presses Universitaires de Rennes, 101-121.

[18] Jean-Marie Homet, 1983, *Astronomes et astronomie en Provence*, Aix-en-Provence, Édisud. Jean-Michel Faidit, 1991, *Les « amateurs » de science d'une province au XVIII^e^ siècle : astronomie et astronomes en Languedoc*, Thèse de doctorat en histoire moderne, Université Montpellier III, par exemple.

[19] Sur les conditions d'acquisition de ces instruments : G. Boistel, 2005, *op. cit.*

Avec la ruée des créanciers des jésuites en 1763, Silvabelle change brusquement de camp et opte pour une attitude rude et inamicale vis-à-vis de ses anciens amis. Pezenas doit se débattre pour ne pas être dépossédé complètement de ses biens et notamment faire reconnaître certains instruments d'astronomie comme lui appartenant, alors que les adversaires des jésuites craignent que ceux-ci ne partent avec la bibliothèque de l'observatoire et déménagent tous les instruments qui font la réputation croissante de cet observatoire depuis les années 1750. Entre les deux hommes, ce ne sont que coups bas et récriminations auprès des autorités locales ou royales. Silvabelle doit s'affirmer comme nouveau directeur de l'observatoire royal de la Marine et ne pas se montrer faible à l'égard des nouveaux ennemis désignés que sont ses anciens amis les jésuites. Pezenas fait tout pour ne pas sombrer avec les autres membres de la Compagnie de Jésus et conserver quelques privilèges et quelques biens lui permettant de poursuivre l'œuvre de sa vie, ce à quoi il parviendra assez bien, grâce aux amitiés et soutiens dont il dispose au plus haut niveau : Académie royale des sciences (Joseph-Nicolas Delisle, Charles-Marie de La Condamine, et dans une moindre mesure, Jérôme Lalande), ministère de la Marine (Machault d'Arnouville, de Boynes) et sans doute le comte de Saint-Florentin, ministre des Cultes[23].

[20] Le mémoire de Silvabelle est examiné à l'Académie des sciences et cause quelques remous dont Joseph-Nicolas Delisle se fait l'écho auprès de Pezenas et du P. Lagrange : « Correspondance de Delisle », Arch. Observatoire de Paris, B1.7, lettres 51, 71 et 78 de février et mars 1753.

[21] Le « Traité des variations célestes ou sur les inégalités du mouvement des planètes » de Silvabelle est publié dans le tome II des *Mémoires de mathématiques et de physique rédigés à l'Observatoire de Marseille*, Avignon, 1756, 201-355.

[22] L'action de Silvabelle à la tête de l'observatoire de Marseille reste à écrire. Ses compétences en astronomie sont souvent mises en cause (par La Condamine lui-même). Les relations entre Silvabelle et le secrétaire perpétuel de l'Académie des sciences de Marseille depuis 1767, et futur maire de cette ville en 1791, Jean Raymond Pierre Mourraille, sont conflictuelles au début des années 1780. L'observatoire est placé sous la tutelle de l'Académie des sciences, lettres et arts de Marseille en 1781 (Arch. de l'observatoire de Marseille, Arch. départementales des Bouches-du-Rhône, 132 J 174).

Dans cette querelle personnelle, Silvabelle parvient à présenter en 1764 un mémoire sur l'utilisation de l'observation des taches du Soleil pour en déterminer la rotation, examiné par Alexis Clairaut et Jérôme Lalande le 15 février 1764 à l'Académie des sciences. Le mémoire (dont on n'a pas conservé l'original) est publié en 1768 dans le tome cinq de la collection dite des « *Savants étrangers* »[24]. L'accueil de ce mémoire est très favorable ; nous donnons la transcription du rapport de Clairaut et Lalande en annexe 3, et le mémoire de Silvabelle en annexe 4. Remarquons que le titre sous lequel est publié le mémoire de Silvabelle – « Problème » – n'est guère engageant et ne laisse rien percevoir de son contenu. Le problème susdit est le suivant : « Trois observations d'une tache du Soleil étant données, déterminer le parallèle du Soleil que décrit la tache & le temps de sa révolution ». On peut voir que Silvabelle expose sa méthode dans le nouveau style mathématique analytique développé par Clairaut, Euler et d'Alembert, notamment, et a de quoi séduire un astronome comme Lalande, ouvert aux nouveautés mathématiques. Remarquons aussi que Silvabelle pose le problème mathématique en le mettant en équation, mais ne fait aucune application numérique ; il ne donne aucune valeur de la période de rotation du Soleil sur lui-même, confirmant ainsi son statut et sa réputation de « Géomètre » que lui confère le rapport académique : Silvabelle, à l'instar de d'Alembert et de Clairaut, déteste le calcul numérique !

---

[23] G. Boistel, 2003, *op. cit.* Voir par exemple la lettre de Pezenas à Charles-Marie de La Condamine, du 26 juillet 1771, d'Avignon, Archives du C.N.A.M., NS5 [PEZENAS]/1.

[24] « *Savants étrangers* » : appellation conventionnelle chez les historiens des sciences pour la collection de l'Académie royale des sciences publiée sous le titre long de : *Mémoires de mathématique et de physique présentés à l'Académie royale des sciences par divers Savans & lus dans ses assemblées* (Paris, Imprimerie royale). Cette collection regroupe des mémoires d'auteurs « étrangers » à l'Académie des sciences, et considérés comme dignes d'intérêt par les académiciens. Elle donne un fantastique aperçu sur l'activité savante au XVIII[e] siècle hors la capitale. G. de Saint-Jacques de Silvabelle, 1768, « Problème », *Mémoires de mathématique et de physique présentés à l'Académie royale des sciences par divers Savans & lus dans ses assemblées*, Paris, Imprimerie royale, 631-634.

Pezenas soumet à l'Académie un mémoire sur les taches solaires qui est examiné le 20 août 1766 par les astronomes Joseph-Nicolas Delisle (correspondant de Pezenas depuis 1750) et le chanoine Alexandre-Guy Pingré. Le problème proposé est le même que celui formulé par Silvabelle : déterminer la période de rotation du Soleil à l'aide de trois observations d'une même tache (position apparente et temps de l'observation pour chacune d'elles). Pezenas l'assortit de plusieurs corollaires et problèmes annexes. Le rapport (reproduit en annexe 2) est favorable ; la méthode de Pezenas, si elle est reconnue ingénieuse, curieuse et utile, ressemble un peu, selon les rapporteurs, à une méthode donnée par Delisle dans les années 1730. Elle est apparentée aux méthodes graphiques développées par Delisle en 1738 dans ses mémoires publiés à Saint-Petersbourg[25] et par Jacques Cassini dans ses *Éléments d'astronomie*, publiés en 1740[26]. À l'aide des observations du P. Martin Poczobut – jésuite polonais ayant séjourné à Marseille entre 1760 et 1762[27] –, Pezenas estime la période de révolution du Soleil à 26 jours 9 heures et l'inclinaison de son axe de rotation sur l'écliptique à 5°14'. Les rapporteurs soulignent que le mémoire de Pezenas est instructif dans le sens où il souligne et corrige des erreurs se trouvant dans les observations anciennes de l'abbé Picard et collectées dans l'*Histoire céleste* de l'astronome Pierre-Charles Le Monnier.

---

[25] Joseph-Nicolas Delisle, 1738, *Mémoires pour servir à l'histoire & au progrès de l'astronomie, de la géographie et de la physique*, Saint-Petersbourg : « Théorie du mouvement des taches du Soleil », 143-179.

[26] Jacques Cassini, 1740, *Éléments d'Astronomie*, Paris, Imprimerie royale : liv. II, chap. II : « De la révolution du Soleil autour de son axe », 86-105.

[27] La présence de Martin Odlanicki Poczobut (1726-1810), s.j., astronome et poète polonais, à l'observatoire de Marseille, est attestée par Silvabelle de juillet 1761 à décembre 1762 au moins. Mais les notes de Silvabelle laissent entendre que Poczobut avait déjà séjourné à Marseille auparavant, au début de l'année 1760 probablement. Il se réfugie à Avignon en 1763 et poursuit ses observations avant de rentrer en Pologne en 1764. Le P. Poczobut deviendra par la suite premier astronome du roi de Pologne, directeur de l'observatoire et recteur de l'Université de Vilnius. Il sera l'un des correspondants de Jérôme Lalande pour l'Académie des sciences. Voir G. Boistel, 2010, *op. cit.*

Le mémoire examiné par l'Académie en 1766 ne sera imprimé qu'en 1774. Il est à peu de choses près identique à celui que Pezenas ajoute à sa traduction du *Cours complet d'optique* de Smith publiée en 1767[28]. C'est aussi la source à laquelle Jérôme Lalande se réfère dans son premier mémoire sur les taches solaires lu à l'Académie en 1775 [29].

Devant les délais de publication des *Savants étrangers* (le tome 5 est publié en 1768 avec le mémoire de Silvabelle ; le tome 6 en 1774 avec le mémoire de Pezenas), Pezenas propose certainement tardivement son manuscrit à la Société royale des sciences de Montpellier[30] qui le refuse car « ayant été imprimé ailleurs » (Figure 1). Compte tenu des diverses dates des rapports et d'impression des recueils des *Savants étrangers*, cette mention de rejet permet d'imaginer que le manuscrit conservé à Montpellier est une réécriture tardive du mémoire soumis à l'Académie royale des sciences en 1766, et a dû être composé après la parution du mémoire de Silvabelle en 1768. Il doit donc dater de la période 1768-72 environ et proposé par Pezenas à la société savante de Montpellier au

[28] E. Pezenas, 1767, *Cours complet d'Optique traduit de l'anglois de Robert Smith*, Avignon, Veuve Girard, Seguin, Aubert, tome 2, « Additions », 524-528. Notons que la traduction du *Cours complet d'optique* de Smith était déjà prête en 1752 comme nous l'avons montré par ailleurs (G. Boistel, 2010, *op. cit.*) ; nous ne savons pas de quand datent les additions du P. Pezenas à sa traduction, mais Pezenas bénéficie entre 1752 et 1755 de remarques des astronomes Nicolas-Louis de Lacaille et Delisle, ainsi que de l'Académicien montpelliérain et médecin, Amoreux. Pezenas et le P. Jean-Baptiste Blanchard s'occupent des mémoires d'optique qu'ils publieront en 1755 dans le tome I des *Mémoires de mathématiques et de physique rédigés à l'observatoire de Marseille.*

[29] Jérôme Lalande, « Mémoire sur les taches du Soleil et sur sa rotation », *Histoire de l'Académie royale des sciences avec les mémoires qui ont été lus pour l'année 1776*, Mém., 457-514, p. 465 en particulier. Lalande signale aussi un paragraphe sur les taches solaires dans l'*Astronomie des Marins* de Pezenas, publiée en 1766. Mais Pezenas ne traite pas des taches solaires en particulier. Il développe par contre des méthodes de trigonométrie sphérique et c'est sans doute à ces méthodes que se réfère Lalande.

[30] Pezenas est élu membre associé de la Société royale de Montpellier et correspondant de Danizy en 1755. Il est en contact avec cette société savante officiellement depuis 1746 (Arch. départ. Hérault, « Registres des délibérations de la Soc. Roy. Sci. de Montpellier », D.120, mars et avril 1755 ; D.121, 6 mars 1755).

moins à la fin de l'année 1774.

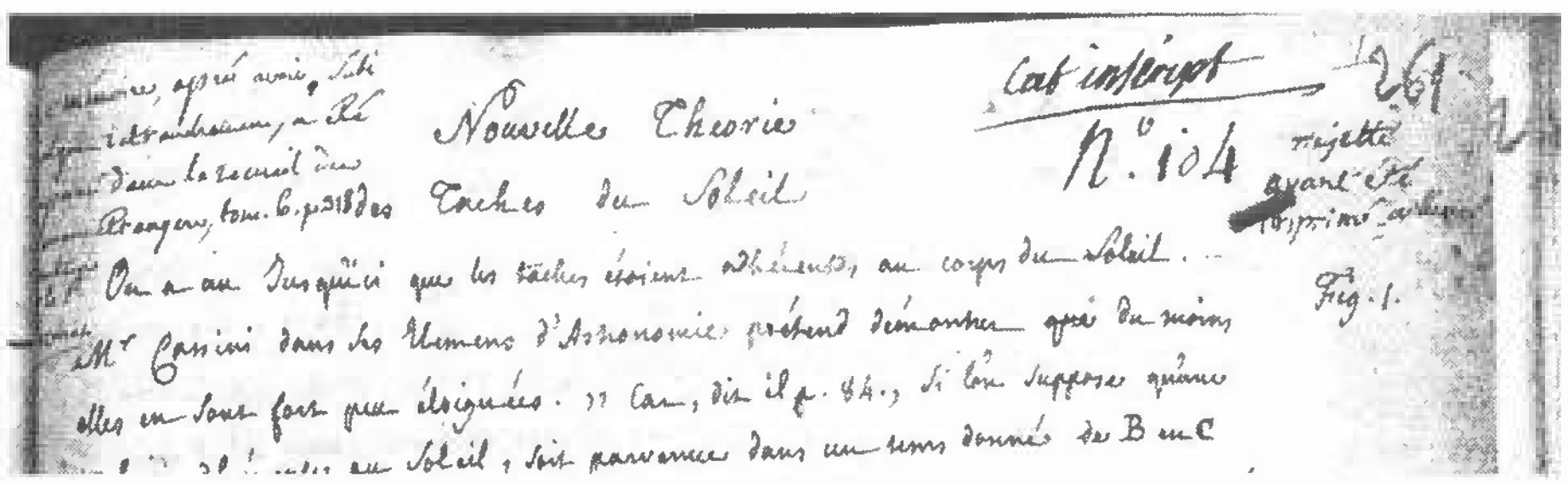

Cat. inscript 261

N°. 104

rejetté ayant été imprimé ailleurs

dans le recueil des Etrangers, tom. 6. p. 318

Nouvelle Theorie des Taches du Soleil

On a cru jusqu'ici que les taches étoient adhérentes au corps du Soleil. ...

Mr Cassini dans ses Elemens d'Astronomie prétend démontrer que du moins elles en sont fort peu éloignées. " Car, dit il p. 84, si l'on suppose qu'une

... au Soleil, soit parvenue dans un tems donné de B en C

Fig. 1.

**Figure 1 : Annotations concernant le rejet du manuscrit de Pezenas par la Société royale des sciences de Montpellier et attribution au P. Pezenas.**

Ce manuscrit montpelliérain comporte une introduction à caractère historique intéressante dans le sens où Pezenas balaye la littérature récente, faisant remarquer qu'il n'est pas si facile pour les astronomes de reconnaître que les taches appartiennent à la surface solaire. Remarquons que c'est une évidence pour le jésuite Pezenas, et que la détermination de la période de rotation solaire s'en trouve ainsi grandement facilitée, à condition de tenir compte du mouvement de la Terre autour du Soleil pendant la durée des observations (entre la première et la troisième observation). Il nous apprend donc que pour certains astronomes de métier au milieu du XVIIIe siècle, l'appartenance des taches à la surface solaire ne va pas encore de soi, même pour un membre de la dynastie Cassini ou un astronome important tel que Georg Wolfgang Krafft de l'Académie Impériale de Saint-Petersbourg. Cette remarque confirme l'*Encyclopédie* qui annonce, à l'article TACHE (du Soleil), que les avis sont très nettement partagés quant à l'adhérence des taches à la surface du Soleil[31].

Le style mathématique que Pezenas emploie est géométrique et graphique en ce sens qu'il pose les relations sous forme de proportions (avec l'usage des signes : ou : :) et que la méthode s'appuie sur les figures géométriques déduites de la position des taches sur le disque solaire. Il est curieux de noter que, malgré ses connaissances et ses traductions d'ouvrages mathématiques comme le *Traité des fluxions* de Colin McLaurin par exemple, Pezenas ne pose pas ces relations sous forme analytique comme il aurait pu le faire et comme l'a fait son adversaire Silvabelle. Pezenas, âgé d'environ 75 ans, est alors dans une période délicate et bouleversée de sa vie. Il doit faire face à la dispersion des jésuites et à une menace très pesante sur ses conditions d'existence.

[31] *Encyclopédie ou Dictionnaire raisonné des sciences, des arts et des métiers, nouvelle édition*, 1779, Vol. 32, Genève, 483-486, 484 en particulier. L'article est en grande partie de Jérôme Lalande.

## Conclusion

Illustrant la pratique d'un astronome d'une longévité peu commune à cette époque, ce texte se trouve à l'articulation de deux traditions savantes et d'une période de transition entre deux styles d'écriture mathématique. Il nous permet de comprendre comment les astronomes du XVIII$^{e}$ siècle, abordaient les problèmes de la détermination des durées de rotation des astres proches de la Terre par diverses méthodes.

# CONVENTIONS TYPOGRAPHIQUES D'ÉDITION

En règle générale, l'orthographe a été légèrement modernisée pour faciliter la lecture, sans entraîner de modification profonde du manuscrit. Ainsi, les accents et certaines conjugaisons ont été retouchés. Le style et la syntaxe sont intégralement préservés.

Entre crochets figurent quelques commentaires de l'éditeur signalant des ratures ou des difficultés de lecture, très peu nombreuses dans ce manuscrit.

Les figures sont les figures d'origine, insérées dans le texte afin de faciliter les renvois et la compréhension des démonstrations du père Pezenas.

La pagination originale du manuscrit conservé aux Archives départementales de l'Hérault, est indiquée comment suit : /261r/ signifie « folio 261 recto » (v pour verso).

Les symboles sont ceux utilisés par le P. Pezenas. En voici la signification :

☉ représente le Soleil.

♈ représente le point vernal.

♎ représente le symbole astronomique/astrologique de la constellation de la Balance.

# REMERCIEMENTS

Au nom des *Cahiers François Viéte*, je tiens à remercier très sincèrement Monsieur Damien Vaisse, Conservateur du patrimoine et adjoint à la directrice aux Archives départementales de l'Hérault, pour son soutien à la publication de ce manuscrit d'Esprit Pezenas.

J'adresse également mes remerciements chaleureux à Mme Colette Le Lay, pour sa relecture attentive, et au Professeur émérite Jacques Gapaillard, pour ses précieuses indications sur la conduite des démonstrations du P. Pezenas, dont certaines ont été intégrées à l'appareil critique de l'édition de ce manuscrit.

*Cahiers François Viète*, Série I, 8, 2004, 21-50.

# « NOUVELLE THEORIE DES TACHES DU SOLEIL »
Esprit PEZENAS

ARCHIVES DEPARTEMENTALES DE L'HERAULT,
MS. D.128, S.D. (C. 1766-70), FOL. 261-267.

EDITION ANNOTEE PAR GUY BOISTEL

/261r/

**Nouvelle théorie des taches du Soleil[1].**

[Fig. 1]

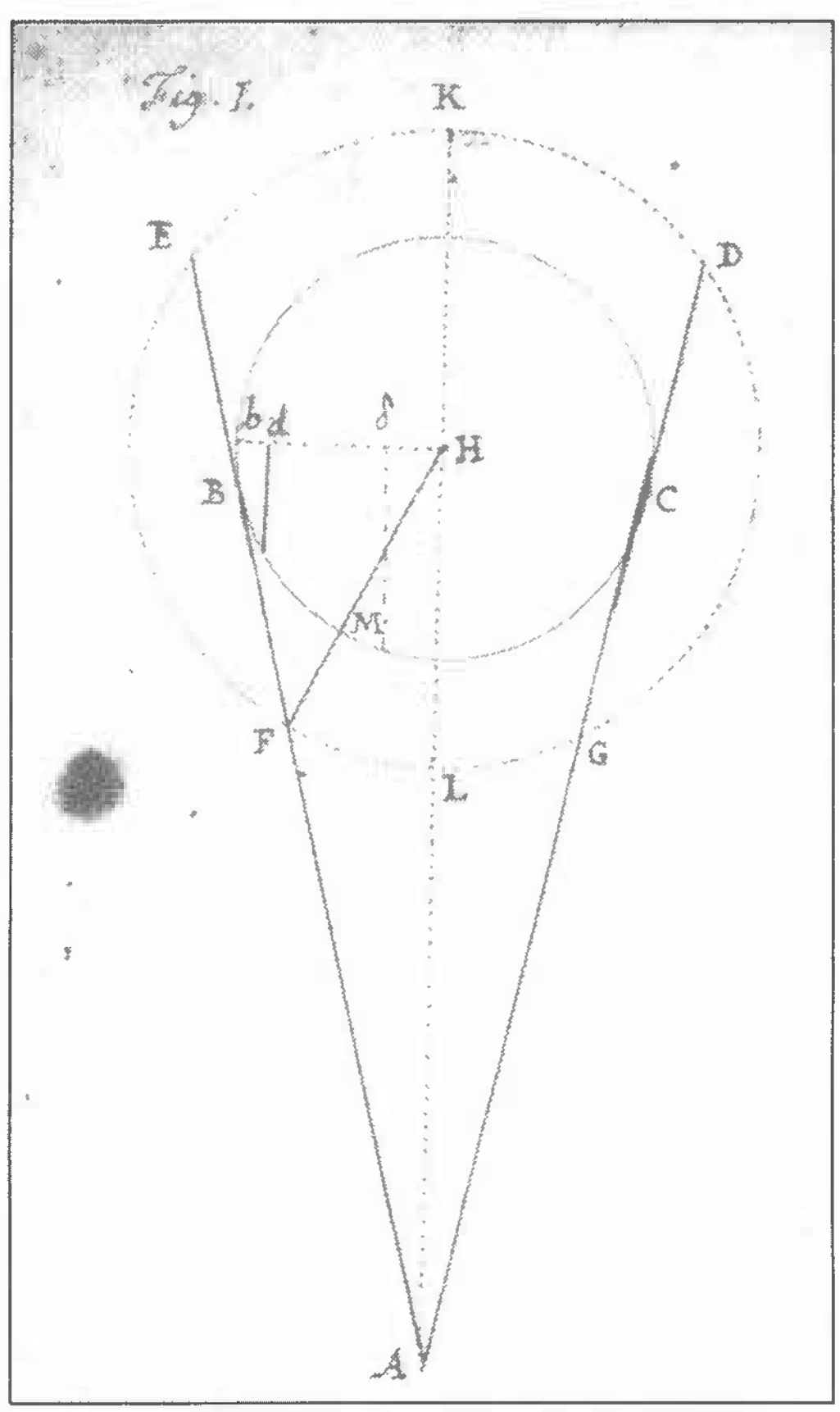

On a cru jusqu'ici que les taches étoient adhérentes au corps du Soleil. Mr Cassini dans les *Élémens d'Astronomie* prétend démontrer que du moins elles en sont fort peu éloignées. « car, dit il p. 84., si l'on suppose qu'une tache adhérente au Soleil, soit parvenue dans un tems donné de B en C [Fig.1] l'espace BC doit être à toute la circonférence qui est de 360 degrés, comme le tems donné est au tems que la tache a employé à retourner au point B ; au lieu que supposant cette tache dans la circonférence du cercle EFGD, parvenue du point F au point G, qui vûs de

[1] Mss ADH, D.128, s.d., fol. 261r – 267v.

la Terre, répondent aux points B et C, l'arc FG qu'elle a décrit effectivement, a un rapport beaucoup plus petit à tout le cercle FGDE, que le tems qu'elle a employé à parcourir l'arc BC, n'a au tems qu'elle emploie à achever toute la révolution. »[2]

Cet argument n'a pas paru fort concluant à Mr. Krafft[3]. Il l'emploie au contraire dans le tome 7 des Mémoires de Petersbourg p. 279 pour prouver que les taches sont éloignées du corps du Soleil d'environ deux demi-diamètres de la Terre[4]. Soit BC le corps du Soleil, et que la tache se meuve dans un cercle concentrique EDGF. Menez de l'œil de l'observateur sur la Terre en A, deux tangentes ACD, ABE au corps du Soleil : il est évident que la tache ne sera visible que dans le tems qu'elle parcourt l'arc FG, si elle n'est pas adhérente au corps du Soleil ; car lorsqu'elle parcourt l'arc GD, ou l'arc FE, on ne sçauroit la voir dans l'atmosphère du Soleil à cause de son obscurité, et lorsqu'elle parcourt l'arc DE, elle nous est cachée par le corps du Soleil. Elle est donc nécessairement invisible pendant tout le tems qu'elle emploie à parcourir l'arc FEDG, et cet arc étant évidemment plus grand que l'arc FG, on voit par ce phénomène que les taches ne sont pas adhérentes à la surface du Soleil.

Si l'on suppose donc qu'elles se meuvent uniformément dans des cercles concentriques autour du Soleil, les arcs FG, FEDG seront proportionnels aux tems ; on pourra donc déterminer ces arcs, si ces tems sont donnés. Menons de la Terre au centre du Soleil la droite AH. Elle divisera en L [tache] et en K les arcs FG, et FEDG en deux parties égales. Donc le tems employé à parcourir l'arc FEK de la moitié de l'occultation, comme l'angle FHL est à l'angle FHK supplément de FHL ; et en composant, la moitié du tems de l'apparition plus la moitié du tems de l'occultation, c'est à dire, la moitié de la révolution totale, est à la moitié du tems de l'apparition, ou ce qui revient au même, le tems de la révolution totale est au tems de l'apparition totale, comme 180 degrés sont à l'angle FHL. On aura donc par le moyen de cette analogie appliquée aux observations, l'angle FHL. Mais /261v/ dans le triangle FAH on connaît la distance AH du Soleil à la Terre, et l'angle FAH ou BAH, qui est le demi-diamètre apparent du Soleil. On aura donc FH et FM.

---

[2] Jacques Cassini, 1740, *Éléments d'astronomie*, Paris, liv. II, chap. I « Des taches du Soleil », 81-105, p. 84 en particulier.

[3] Il s'agit de Georg Wolfgang Krafft (1701-1754), le père de l'astronome Wolfgang Ludwig Krafft (1743-1814).

[4] Georg Wolfgang Krafft, « De invenienda distantia macularum solarium a Sole », *Commentarii Academiae scientiarum imperialis Petropolitanae pro Annis 1734 & 1735,* Saint-Petersbourg, 1740, 279-282 (+ pl.)

L'auteur applique ensuite sa méthode à une observation célèbre faite à Leipsic (sic) par Godefroi Kirchius[5] en 1684. Il vit revenir plusieurs fois une même tache depuis le 26 avril jusques au 7 juillet, et il en conclut que le tems de son apparition étoit de 12 jours, et que celui de son occultation étoit de 15 jours[6]. Pour achever le calcul, Mr Krafft suppose avec Mr. de la Hire[7], que la distance moyenne de la Terre au Soleil est de 34377 demi-diamètres de la Terre[8], et que le demi-diamètre moyen apparent du Soleil est 16'. 15", et il fait cette proportion : comme le rayon est à la distance 34377 du Soleil à la Terre, ainsi le demi-diamètre apparent du Soleil est à son demi-diamètre réel (160,829 demi-diamètres de la Terre). Notre auteur fait ensuite cette proportion : comme le tems de la révolution totale (27 jours) est au tems de l'apparition (12 jours) ainsi 180 degrés sont à l'angle FHL, qu'on trouve de 80 degrés[9]. Dans le tems de l'observation, le demi-diamètre apparent étoit de 15'.54", et le logarithme de la distance du Soleil à la Terre étoit de 4.00504, en supposant la distance moyenne 10000. Le nombre qui répond à ce logarithme est 10116,74. Donc en réduisant cette distance en demi-diamètres de la Terre, et supposant la distance moyenne 34377 demi-diamètres, la distance du Soleil à la Terre étoit alors 34778,317 demi-diamètres de la Terre. Donc par la résolution du triangle FHA, dont on a deux angles, FAH = 15'.54", FHA = 80°. 0'. 0", et le côté AH = 34778,317 ; on trouvera FH distance de la tache au centre du Soleil = 163,204 demi-diamètres de la Terre ; d'où soustrayant le demi-diamètre trouvé du Soleil, on aura la distance de la tache à la surface du Soleil 2,375 demi-diamètres de la Terre.

---

[5] Gottfried Kirch, écrit aussi Kirche ou Kirchius (1639-1710), père de l'astronome Christfried Kirch (1694-1740). Ils furent tous deux directeurs de l'observatoire de Berlin. Gottfried Kirch est le premier à avoir observé la comète de 1680. Il forma sa seconde femme, Maria Margueritte née Winkelman (1670-1720) et ses deux filles à calculer les éphémérides astronomiques de Berlin (J. Lalande, 1802, *Bibliographie astronomique*, Paris, 286-88, 298, 898). J.-N. Delisle fit l'acquisition d'une grande partie de la correspondance et des observations de Kirch qui furent remises au Dépôt de la Marine à Paris. Une partie de ces archives se trouvent désormais à la bibliothèque de l'observatoire de Paris (voir la base de données Alidade, au nom de Godefroy Kirch).

[6] Les sources divergent un peu quant aux données de cette observation de Kirchius. Cette observation est donnée pour l'année 1684 par Christian Wolff (Chretien Wolf) dans son *Cours de mathématiques, qui contient toutes les parties de cette science mises à la portée des Commençans (l'optique, la catoptrique, la dioptrique, la perspective, la Géographie, la chronologie, la Gnomonique, l'Astronomie & la Navigation)*, 1747, tome 2, Paris, Charles-Antoine Jombert, page 253 [l'ouvrage ne peut avoir échappé au P. Pezenas, compte tenu de ses lectures et de ses relations avec Jombert], depuis le 6 avril jusqu'au 17 juin, tache observée simultanément par Jean Dominique Cassini à l'Observatoire royal de Paris (*Journal des sçavans*, 1684, pp. 177-180 (+ pl.), 238-240). L'observation est indiquée dans l'*Encyclopédie, op. cit.*, 1779, Vol. 32, 484 ; observation d'une tache solaire par Kirchius, datée de 1681 [erreur puisque attestée dans le *Journal des sçavans* pour 1684] depuis le 26 avril jusqu'au 17 juin. Quoi qu'il en soit, la tache a été visible pendant douze jours sur le limbe et invisible pendant 15 jours, ce qui conduit à une période de 27 jours pour la rotation du Soleil. Les observations de Cassini et de Kirchius notamment ont conduit certains astronomes à douter que les taches appartenaient à la surface solaire (*Nouvelles vuës sur le système de l'Univers*, 1751, Paris, Chaubert et Ballard, p. 154). Cette observation « célèbre » est encore mentionnée par la Société philomatique de Paris, dans son *Nouveau bulletin des sciences*, en 1826, p. 66 : les durées de visibilité et d'invisibilité des taches ne sont pas égales, ce que confirme un correspondant de la Société, M. Emmett, qui trouve une durée de visibilité de 12 jours 8 heures 30 minutes et une durée d'invisibilité de 15 jours 3 heures 30 minutes.

Cette démonstration suppose : 1°. que la Terre a été immobile pendant tout le tems de la révolution de la tache, 2°. que cette tache a fait sa révolution dans son équateur, qui seul est concentrique au Soleil, 3°. que l'observation de Kirchius est incontestable : trois hypothèses qu'on ne peut pas admettre.

Je dis en premier lieu que cette démonstration suppose que la Terre a été immobile pendant tout le tems de la révolution de la tache. En effet, notre Auteur suppose que dans 27 jours la tache n'a parcouru que 360 degrés, tandis que le mouvement moyen de la Terre lui a fait parcourir 26°.36'.36" de plus. Car la tache ayant paru entrer en B dans le disque du Soleil, et ce point du disque ayant parcouru avec la Terre 26°.36'.36" dans 27 jours, l'observateur aura vu r'entrer (sic) la tache dans le disque après qu'elle aura décrit 360° plus 26°36'36". Donc il faut changer la proportion en cette manière : 27 jours sont à 12, comme 180° + 13°.18'.18" sont à 85°,91 = FHL ; et le triangle FHA donnera alors AFH = 93°.49'.30" et FH = 161,21 demi-diamètres de la Terre, et le demi-diamètre réel du Soleil étant = 160,83, la distance de la tache à sa surface ne sera que 0,38.

---

[7] Philippe de la Hire (1640-1718). Troisième réédition de ses tables astronomiques, *Tables du Soleil et de la Lune* (1687) en 1735. Son fils Gabriel-Philippe de la Hire (1677-1719) poursuivit les mesures des diamètres solaires initiées par son père.

[8] En supposant un rayon équatorial terrestre égal à 6378 km en moyenne, la distance Terre-Soleil est évaluée à cette époque à 34377 fois 6378 km, soit 219 256 506 km, distance très largement surévaluée de près de 70 millions de kilomètres, qui illustre bien les progrès que doit encore faire la mécanique céleste au milieu du XVIII$^{e}$ siècle. En outre, les 34377 rayons terrestres de La Hire correspondent à une parallaxe solaire de 6" qui sont une exception au milieu du XVIII$^{e}$ siècle où la tendance générale est plutôt de surestimer cette parallaxe autour de 10" ou 12".

[9] Lire comme suit : $\frac{27}{12} = \frac{180°}{FHL}$ soit $FHL = \frac{180° \times 12}{27} = 80°$.

L'Auteur suppose en second lieu que cette tache a décrit pendant son apparition un arc FG concentrique au Soleil, et par conséquent un arc de grand cercle, [sur] une partie de l'Équateur de sa révolution [?]. On voit au contraire par les observations de Mr. Picard[10] publiées par Mr. Le Mon[n]ier dans son *Histoire Céleste*[11], que la route de cette tache n'a pas paru traverser le Soleil avec 12°. de déclinaison. Si nous donnons cette déclinaison à la tache de 1684, nous dirons : le rayon est au cosinus de 12°., comme le demi-diamètre apparent du Soleil (15'.54") est au demi-diamètre apparent du parallèle de la tache, que l'on trouvera de 15'.33" ; et l'angle BFH étant alors de 86°. 6'. 3", on dira : comme le sinus de 86°. 6'. 3" est à AH = 34778,317, ainsi le sinus de 15'.33" est à FH, que l'on trouvera = 157,68[12] ; et le demi-diamètre réel du Soleil étant au demi-diamètre de ce parallèle, comme le rayon est au cosinus de 12°, on trouvera

---

[10] L'abbé Jean Picard (1620-1682), astronome du roi et auteur de nombreux travaux en géodésie (réforme de la carte du royaume de France) entre autres. L'un des 21 premiers membres de l'Académie royale des sciences. Il a initié de vastes séries d'observations systématiques planétaires et stellaires en astronomie.

[11] L'*Histoire céleste ou recueil de toutes les observations astronomiques faites par ordre du Roy* (Paris, Briasson), de l'astronome Pierre-Charles Le Monnier (1715-1799), publiée en 1741, est un vaste recueil d'observations astronomiques depuis le XVII[e] siècle, observations astronomiques diverses et suffisamment précises pour être exploitées par les astronomes pour des études à mener sur le long terme : éclipses, taches du Soleil, observations de comètes, etc. Les observations de la tache correspondante à celle de Kirchius et effectuées par Picard, sont publiées par Le Monnier, pp. 314 à 325.

[12] Dans une écriture plus courante :

$$FH = \frac{\sin(15'33'') \times 34778{,}317}{\sin(86°6'33'')} = \frac{0{,}0045 \times 34778{,}317}{0{,}9977} \approx 157{,}68.$$

que le demi-diamètre du parallèle de la tache dans le Soleil est de 157,31 demi-diamètres de la Terre, ce qui étant soustrait de FH = 157,68, donnera la distance de la tache à la surface du Soleil dans la direction de son parallèle = 0,37. Ce qui revient au premier calcul.

Dans la 3e hypothèse on prétend que Kirchius a observé fort exactement le tems de l'apparition de cette tache. Il faut pour cela qu'il ait observé l'entrée de la tache dans le disque et le mouvement de sa sortie. Or ces deux observations sont très difficiles à moins qu'on n'ait un télescope qui grossisse plus de cent fois[13]. Car si l'on suppose que la tache parcourt 15°. par jour, l'angle visuel sous lequel on voit cet intervalle auprès du centre, est à celui sous lequel on le voit auprès de la circonférence, comme le ~~co~~sinus Hδ de 15° est au sinus verse bd du même nombre de degrés à fort peu près [voir fig.1.]. Mais le ~~co~~sinus de 15° est à son sinus verse, comme 1 est à 0,132 [rature]~~0,~~ et l'intervalle d'un jour paroit auprès du centre du Soleil sous un angle d'environ deux [rature] minutes ; donc le même intervalle auprès de la circonférence ne doit paroître que sous un angle de 0',264 [ratures], c'est à dire sous un angle de [ratures] 16 secondes. Je doute qu'on puisse distinguer un intervalle de 2 secondes avec un télescope qui ne grossiroit pas cent fois. Or [ratures] comme la grandeur apparente de la tache auprès [des] bords diminue aussi dans la même proportion, il faudroit s'assurer de l'intervalle de [ratures] deux secondes, pour pouvoir déterminer à un demi-jour près le moment de l'entrée d'une tache dans le disque du Soleil et le moment de sa sortie.

Mr. Cassini dans ses *Élémens d'Astronomie* réduit à 5 Problèmes toutes les méthodes connues pour déterminer la révolution du Soleil autour de son axe. Dans le premier, il donne la manière de déterminer sur le disque apparent du Soleil, la situation du parallèle qu'il décrit par rapport à l'Écliptique. /262v/ Dans le second, il détermine dans le même disque la situation des taches par rapport à l'Écliptique. Ces deux problèmes ne souffrent aucune difficulté. Dans le troisième, il donne la méthode de déterminer par les observations des taches, la situation du pôle de la révolution du Soleil autour de son axe, et l'inclinaison de cet axe à l'égard de l'Écliptique. Cette méthode suppose que les taches paroissent décrire sur le disque du Soleil une ligne qui sera tantôt droite, tantôt elliptique. La ligne perpendiculaire [insertion : au milieu] de cette ligne droite, ou du grand axe de l'ellipse, passera par le centre du disque et formera avec l'axe de l'Écliptique un angle qui mesurera l'inclinaison de l'axe de rotation par rapport à l'Écliptique. Cette supposition seroit évidente, si le disque du Soleil étoit toujours le même. Mais comme il varie tous les jours et qu'il suit le mouvement de la Terre autour du Soleil, la projection ne se fait pas sur le même plan. Le méridien étant toujours représenté par une ligne droite, et l'Écliptique ne formant pas toujours le même angle avec le méridien, il est évident qu'on ne peut pas la représenter par une ligne droite, comme on le fait dans cette projection.

[13] Le P. Pezenas peut faire valoir les grossissements (200 ou 800 fois) dont il dispose avec les nouveaux télescopes de Short (cf. G. Boistel, 2005, *op. cit.*).

Dans le problème 4$^{e}$, Mr. Cassini détermine pour tous les jours de l'année la situation apparente du pôle de la révolution du Soleil et les ellipses que les taches paroissent décrire ; et dans le 5$^{e}$., il détermine le tems de la révolution des taches autour de leur axe. Il faut pour cela déterminer la situation du pôle boréal du Soleil dans le tems qu'une tache est vers le milieu de son cours apparent dans le disque, et ayant mené par ce pôle un diamètre qui représente un cercle de déclinaison, on observera le tems vrai auquel la tache passe par ce cercle, et pour une plus grande exactitude, on réduira ce tems vrai en tems moyen. On observera ensuite le tems moyen auquel la même tache après avoir fait une révolution entière, revient au cercle de déclinaison qui passe par le pôle du Soleil et par son centre ; ce cercle ayant changé de situation apparente sur le disque. L'intervalle entre ces tems mesure la révolution apparente du globe du Soleil à l'égard de la Terre. Ensuite pour réduire cette révolution observée en tems moyen à la révolution véritable, on dira : comme 360°. plus le mouvement vrai du Soleil dans l'intervalle de la révolution observée, sont à 360°. plus le mouvement moyen qui convient à cet intervalle ; ainsi l'intervalle entre les tems de la révolution observée, qu'on a réduit en tems moyen, est au tems de la révolution véritable, que l'on a déterminé par un grand nombre d'observations, de 27 jours 12 heures 20 minutes[14].

---

[14] Il n'est pas très clair de retrouver dans ce qui précède ce que l'auteur a voulu dire (et qui correspond à ce qui suit) en combinant, à juste titre, la rotation du Soleil et la révolution annuelle de la Terre. La « révolution apparente » ou « observée » « du globe du Soleil » est sa rotation synodique. Si l'on néglige l'inclinaison de l'équateur solaire sur l'écliptique, la vitesse angulaire de cette rotation synodique est la somme de la vitesse angulaire de la rotation sidérale du Soleil et de la vitesse angulaire de la révolution sidérale de la Terre. Il en résulte, si $T$, $T'$ et $T''$ sont respectivement (les durées de) ces rotations et révolution, la formule $\frac{1}{T} = \frac{1}{T'} + \frac{1}{T''}$, d'où l'on peut déduire $T$ de $T' = 27$j 12h 20m et de $T'' \approx 365{,}256$j. C'est cette même formule qui établit la correspondance indiquée

Lorsqu'une tache ne passe pas assez de tems sur la surface du Soleil, pour décrire une révolution entière, on déterminera sa situation en divers jours, comme D, E et F (fig.2.) et l'on tracera l'ellipse AEB de sa révolution apparente. Sur le grand axe AB* de cette ellipse on décrira le demi-cercle AGB, et des points D, E, F on mènera les lignes DH, EG, FI perpendiculaires à AB. Les arcs HG, GJ mesureront sur la circonférence AGB les arcs diurnes de la révolution du Soleil, et l'on fera : comme l'arc HI est à 360° /263r/ ainsi le nombre de jours et d'heures que la tache a employé à parvenir de D en F, est à toute la révolution de la tache autour du Soleil.

*Note* [du P. Pezenas] */bas de page 262v/ *. AB ne peut pas être le grand axe de l'ellipse ; parce que cette ellipse ne peut pas toucher intérieurement le bord du Soleil, ou avoir une tangente commune au cercle et au sommet A de l'ellipse. Car la tangente au cercle est perpendiculaire au rayon CA, et la tangente au sommet de l'ellipse est perpendiculaire au grand axe.*

[Fig. 2.]

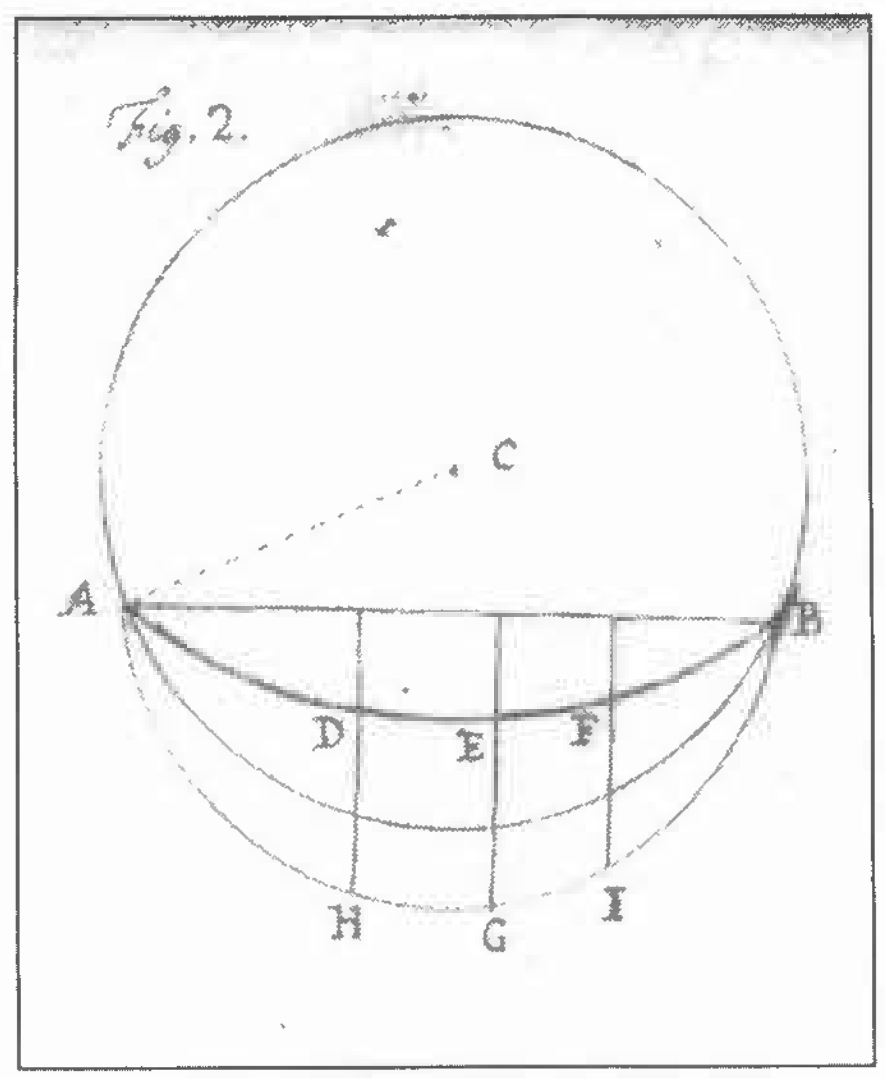

par A. Danjon (*Astronomie générale*, *op. cit.*, p. 350) entre la rotation sidérale du Soleil de 25,38j et sa rotation synodique de 27,275j.

Enfin pour déterminer cette révolution à l'égard d'un point fixe dans le Ciel, on dira : comme 360°. plus le mouvement vrai de la Terre pendant la révolution apparente, sont à 360°. plus le tems moyen que la tache a employé à retourner au cercle de latitude qui passe par le centre du Soleil et de la Terre, est à la révolution de la tache à l'égard d'un point fixe dans le Ciel. Cette dernière proportion seroit très exacte, si la tache avoit fait sa révolution autour de l'axe de l'Écliptique. Mais son axe de rotation étant incliné à l'axe de l'Écliptique, (fig.3), p le pôle de la rotation : si la tache a paru en T dans le cercle de latitude ΠT, lorsque la Terre étoit en Z, la Terre ne la verra plus en T, mais en T' dans le cercle de latitude ΠT', lorsqu'elle sera en Z' et que la tache serea revenue au milieu du Soleil : de sorte que la tache aura parcouru 360°. plus l'arc TT' ou l'angle TpT', la Terre n'ayant parcouru que l'angle TΠT' dans cet intervalle. Or cet angle est plus petit que l'angle TpT' lorsque le pôle p est entre Π et T, et il est plus grand lorsque Π est entre p et T. Donc le premier terme de la proportion étant trop petit ou trop grand, le dernier terme sera nécessairement trop grand ou trop petit.

[Fig.3.]

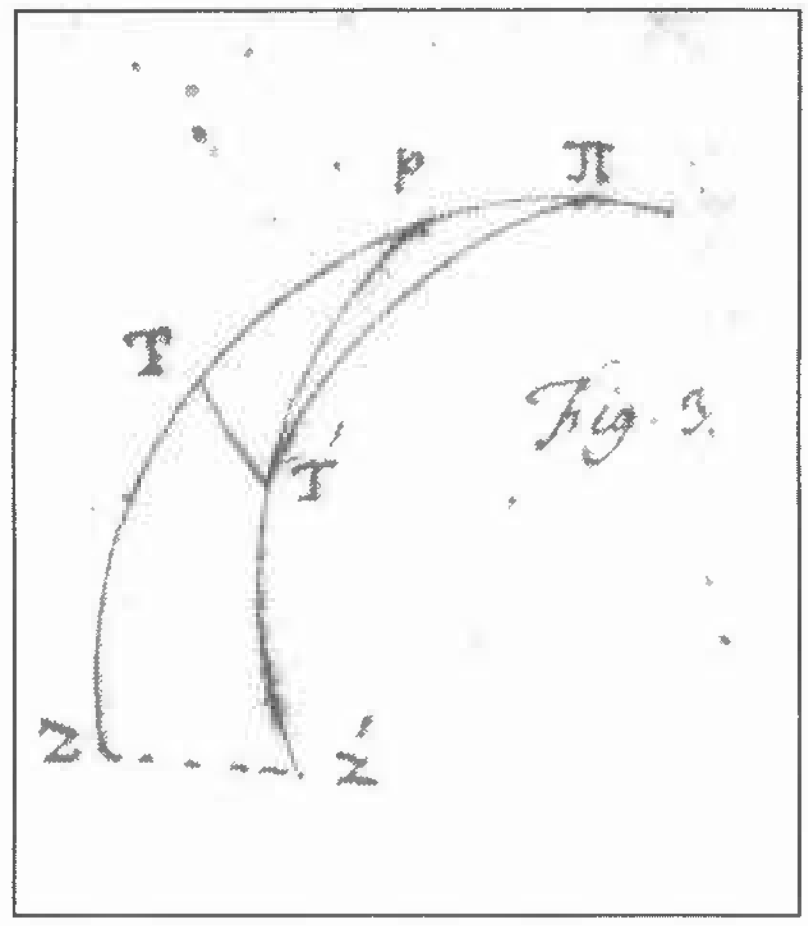

Pour éviter tous ces inconvénients, nous avons employé la méthode suivante, qui se réduit à ce problème général : trois observations étant données d'une même tache en trois tems différents, trouver le tems de sa révolution, et la position de son axe, ayant égard au mouvement de la Terre, et en considérant la tache comme se mouvant sur la surface du globe immobile du Soleil.

Pour résoudre ce problème plus aisément, nous le divisons en plusieurs autres forts (sic) simples.

Définitions[15].

J'appelle Zenith du Soleil le point Z (fig. 4.) le plus proche du centre de la Terre, ou de l'œil O de l'observateur. Ce point est toujours dans la ligne qui joint les deux centres C et O, et cette ligne est toute entière dans l'Écliptique. AZB est le demi-cercle du Globe du Soleil qui est dans le plan d'un parallèle à l'Équateur ou d'un almicantarath ; CA son demi-diamètre perpendiculaire au rayon visuel OC. OD est un autre rayon visuel qui touche le quart de cercle AZ en D, et qui est par conséquent perpendiculaire au demi-diamètre CD. L'arc DA qui mesure l'angle DCA complément de ZCD, est aussi la mesure de demi-diamètre apparent COD, complément de ZCD.

Si l'on imagine un petit cercle dont le diamètre soit DG parallèle à AB, et qui soit perpendiculaire au plan BZA, ce petit cercle représentera le disque du Soleil : son centre F sera la projection du Zenith Z ; les lignes droites DF, FG seront les projections des arcs DZ, GZ ; et par la même raison tous les demi-diamètres du disque du Soleil, seront les projections de tous les arcs qui se coupent en Z.

/263v/

[15] Ce fragment du manuscrit correspond en grande partie à ce qui a été imprimé dans le *Cours complet d'optique* de R. Smith traduit par Pezenas et publié en 1767 (cf. *infra*) et dans les mémoires des « savants étrangers » en 1774 reproduit en annexe 1. On notera un profond remaniement du texte original de Pezenas dans ce dernier mémoire.

[Fig. 4.][16]

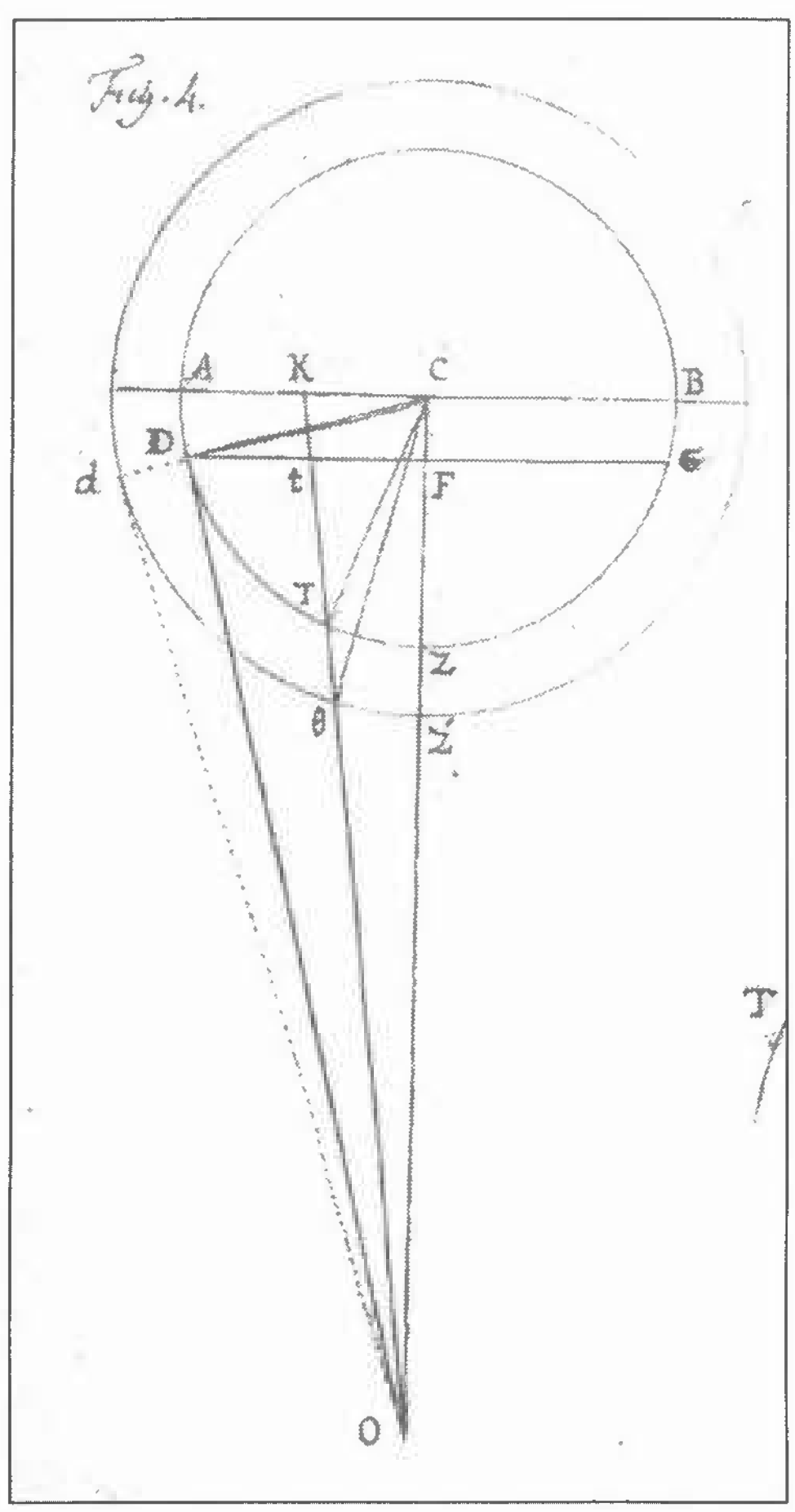

Problème 1.

Trouver la grandeur de l'arc ZT compris entre le Zénith Z et une tache T quelconque, dont on a la distance apparente COT, et que l'on suppose placée sur la surface du Soleil ?

[16] Remarque : la figure 4 n'est pas tout à fait correcte, car, comme on le lit dans la suite, les rayons CD et Cd étant respectivement perpendiculaires aux tangentes OD et Od, les points C, D et d ne peuvent être alignés.

Prolongez OT jusqu'au diamètre réel en K ; vous aurez ces proportions : le sinus de COT est au sinus de CTO ou de CTK : : CT ou CD : CO : : (à cause des triangles rectangles CDO et CDF qui ont un angle commun FCD) CF : CD : : le sinus de l'arc DA est au rayon. Donc le sinus du demi-diamètre apparent, ou de l'arc DA, est au rayon, comme le sinus de la distance apparente de la tache au Zénith, est au sinus de l'angle CTK extérieur au triangle COT. Donc en retranchant de cet angle CTK l'intérieur COT, on aura l'angle ZCT, ou l'arc ZT. L'observation donne le demi-diamètre apparent COD et la distance apparente COT de la tache T au Zénith, soit qu'elle se trouve dans l'arc ZTD, ou dans un autre grand cercle du globe du Soleil qui passe par le point Z. On aura donc l'angle CTK, qui est toujours plus grand que l'angle ZCT. Si l'œil étoit à une distance infinie du Soleil, la projection Ft de ZT seroit égale au sinus de cet arc. Mais on ne peut pas admettre cette hypothèse, et Ft est toujours plus grand que le sinus de ZT sous le rayon CT.

Si la tache étoit à quelque distance de la surface du Soleil comme en θ, il faudroit prendre le demi-diamètre apparent COd du globe dont le rayon est Cθ, et la projection étant la même, on aurait l'angle Cθt plus petit que CTK, et Z'θ plus petit que ZT.

## Corollaire

Cette méthode s'applique naturellement aux satellites de Jupiter ou de Saturne. Soit θ l'un de ces satellites, COd sa plus grande digression. Si l'on suppose que l'orbe du satellite soit circulaire, on dira : comme le sinus de la plus grande digression COd, est au rayon, ainsi le sinus de la distance apparente du satellite θOZ', est au sinus de l'angle CθK. ce qui donne l'arc Z'θ.

[fig. 5.]

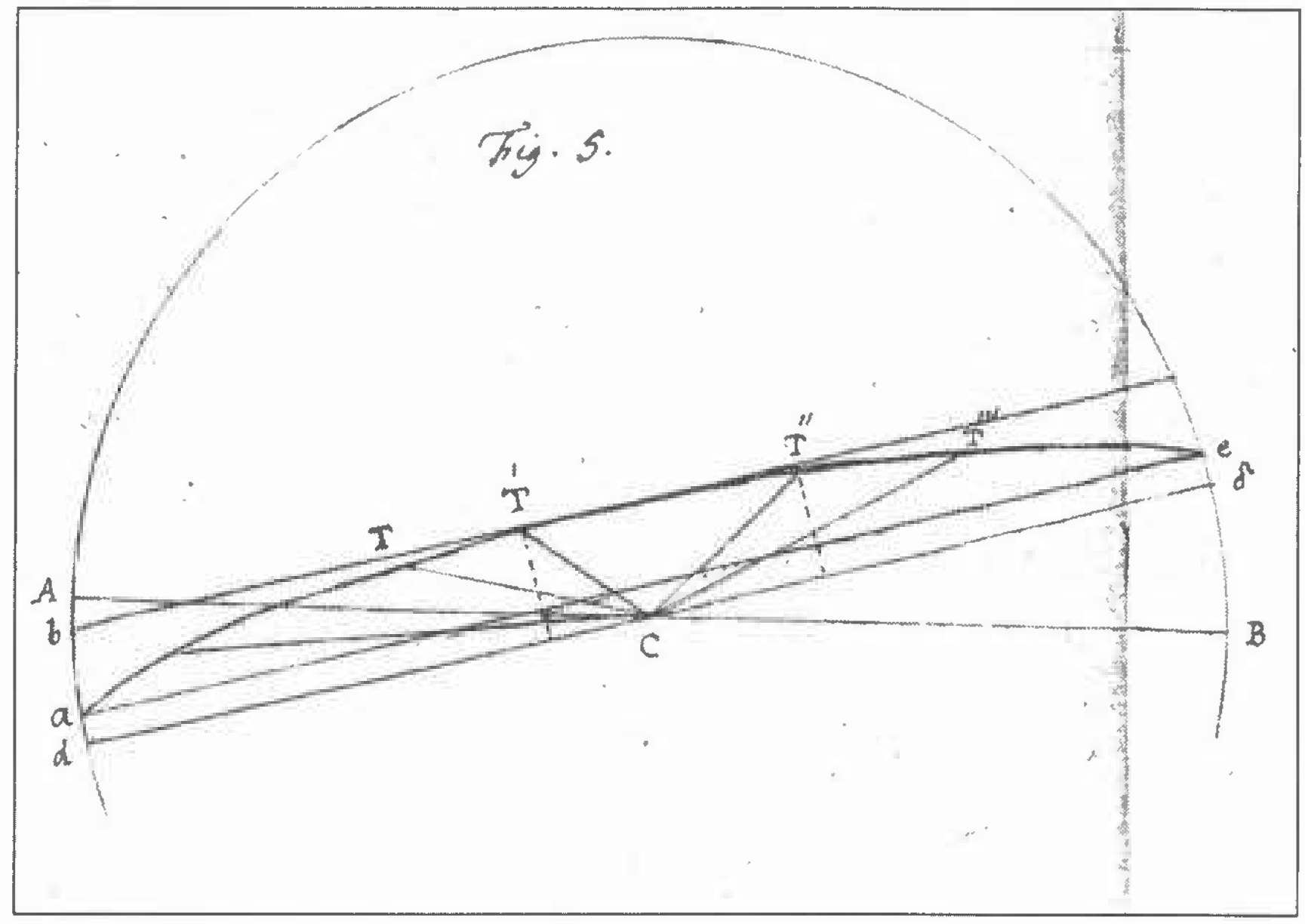

Problème 2.

Trouver la valeur de l'angle formé au Zénith par le grand cercle ZD et par l'arc ZT ?

[fig. 4. et 5.]

L'angle rectiligne TCA (fig. 5.) formé au centre du disque par la distance TC de la tache à ce centre, et par le demi-diamètre horizontal CA, est égal à l'angle sphérique formé par l'arc de grand cercle ZT (fig. 4.) dont la projection est CT, et par l'arc ZD dont la projection est CA. Car ces deux angles sont mesurés par le même arc du cercle de projection dont C est le centre et dont le pôle est le Zénith. Or les observations donnent les lignes TC et CA, et l'angle TCA. Donc on aura par observation l'angle sphérique requis[17].

[17] Commentaire sur la figure 4 : sur la surface sphérique du Soleil, l'angle des arcs de grands cercles ZT et ZD est l'angle de leurs tangentes en Z (appelons-les Zt' et Zd'). Or, dans la projection de centre O sur le disque du Soleil, comme le

[fig. 6.]

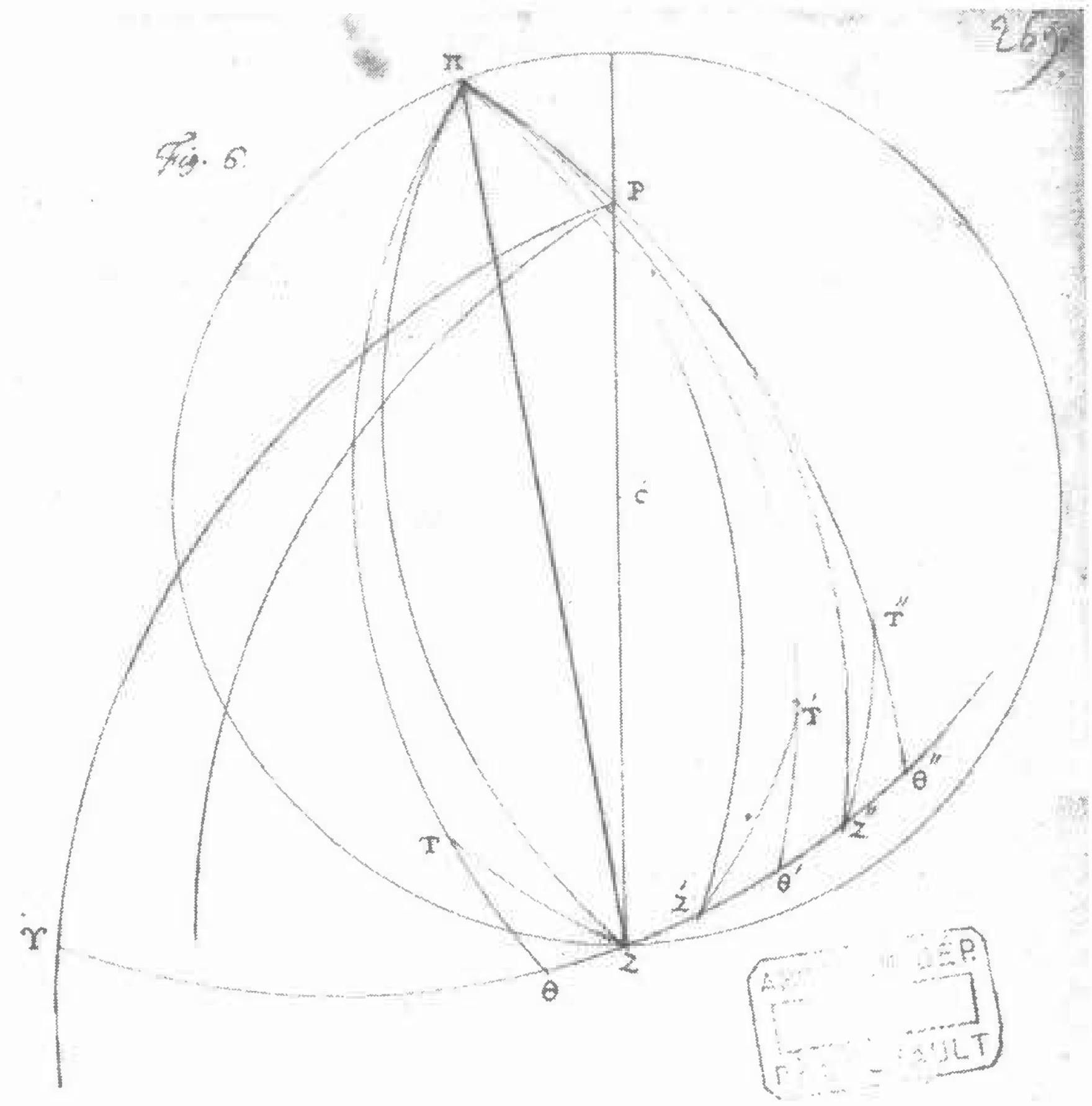

Problème 3.

Trouver la latitude et la longitude d'une tache du Soleil ?

---

plan de ce disque et le plan t'Zd' sont parallèles (puisque tous deux perpendiculaires à la droite OZC), l'angle t'Zd' (= angle sphérique TZD) se projette suivant l'angle TCA qui lui est égal. Notons que le P. Pezenas considère que la projection de D est A, ce qui n'est pas exact puisque les points O, D et A (fig. 4) ne sont pas alignés, sans que ceci ait une incidence sur son résultat d'égalité d'angles.

Soit ♈ le plus prochain Équinoxe dans le globe du Soleil (fig. 6.) ♈Z l'Écliptique et la distance de la Terre à un Équinoxe. P le pôle de l'Équateur dans le méridien PCZ. On a les trois cotés du triangle ♈ZP, sçavoir ♈P de 90°., la distance ZP du lieu /264r/ de la Terre au pôle Nord, et la distance ♈Z au plus prochain Équinoxe. On aura donc l'angle ♈ZP. On auroit également cet angle par le moyen de l'ascension droite ♈PZ, en disant, sin. ♈Z : sin. ♈PZ : : sin. ♈P = rayon : sin. ♈ZP. Soustrayant ensuite de cet angle ♈ZP, l'angle TZP dont on a trouvé le complément TZD par le probl. 2., ou ajoutant T'Z'P, si la tache est plus occidentale que ZP, on aura l'angle ♈ZT ou ♈Z'T'.

Soit du pôle Π de l'Écliptique le cercle de latitude ΠTθ perpendiculaire à l'Écliptique en θ. Dans le triangle rectangle TZθ, on connoit TZ et l'angle TZθ; on aura donc la latitude Tθ de la tache, et sa distance TΠ au pôle de l'Écliptique ; et dans le triangle TZΠ, connoissant les trois cotés TZ, TΠ, et ZΠ de 90°, on trouvera TΠZ différence en longitude entre la tache T et la Terre Z[18].

---

[18] Il faut comprendre que Pezenas fait ici de la trigonométrie sphérique sur la surface du Soleil dont le disque est représenté sur la figure 6. Pour cela il adopte un point de vue héliocentrique où la sphère solaire, que nous nommerons Σ, se substitue à la sphère céleste géocentrique habituelle. Par projection de centre C (centre du Soleil) sur Σ, la figure 6 est censée représenter en particulier : 1°. la position Z de la Terre lors de l'observation de la tache en T ; 2°. la trace sur Σ du plan de l'écliptique, laquelle doit donc passer par Z et par la projection γ (A dans le texte) sur Σ de la position de la Terre à l'équinoxe de printemps. (Dans la figure 6, il semble que Pezenas confonde la position γ de la Terre à l'équinoxe et sa projection A sur Σ, et il en résulte une figure 6 « surréaliste » où γ = A est extérieur à Σ !) ; le pôle céleste P (tel que CP soit parallèle à l'axe du monde) ; le pôle Π de l'écliptique (tel que CΠ soit perpendiculaire au plan de l'écliptique). L'objet de la démonstration est de déterminer les latitude et longitude de T sur Σ, ces coordonnées sur Σ étant comptées par rapport au plan de l'écliptique et au grand cercle passant par Π et A. Tous les triangles considérés ci-dessous sont des triangles sphériques sur Σ. Dans le triangle rectilatère AZP (AP = 90° car, aux équinoxes, le plan équatorial passe par A et C), les côtés ZP (= 90° + déclinaison du Soleil) et AZ (= longitude du Soleil) sont supposés connus à l'époque de l'observation, et l'angle AZP s'en déduit par : cos AZP = – cotan ZP.cotan AZ.

Ou encore, en utilisant l'angle APZ (= ascension droite du Soleil, elle aussi supposée connue), on a encore dans le triangle rectilatère AZP :

Problème 4.

Les latitudes et longitudes d'une même tache étant données en deux tems différents, trouver l'arc de grand cercle qui joint ces deux positions ?

Le Zénith Z se mouvant avec la Terre, et Z étant le lieu de la Terre dans la première observation, Z' sera le lieu de la Terre dans la seconde, si l'arc ZZ' est égal à l'accroissement de la longitude de la Terre entre les deux observations. On trouvera donc comme dans le Probl. 3. l'arc $\Pi T'$, et l'angle $Z'\Pi T'$, qui étant combiné avec les angles $T\Pi Z$, $Z\Pi Z'$, donnera l'angle $T\Pi T'$. Donc connoissant les deux cotés $T\Pi$, $T'\Pi$ et l'angle compris $T\Pi T'$, on aura la base TT' du triangle $T\Pi T'$ [19].

[fig. 7.]

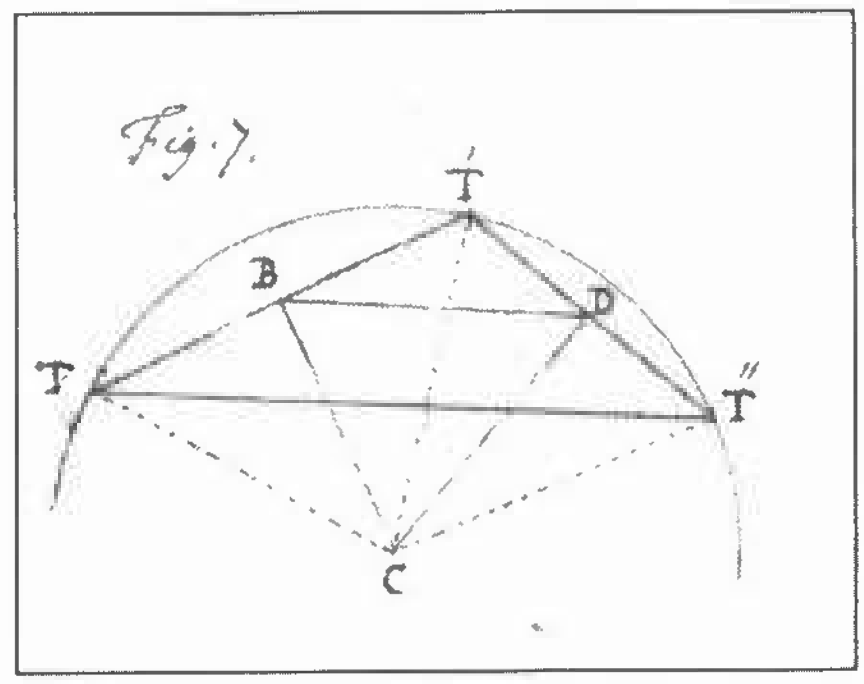

Problème 5.

Les latitudes et longitudes d'une même tache étant données en trois tems

---

$\sin AZP = \frac{\sin APZ}{\sin AZ}$. Puis, dans le triangle $T\theta Z$, rectangle en $\theta$, on obtient la latitude $T\theta$ par : $\sin T\theta = \sin ZT.\sin \theta ZT$, d'où l'arc $T\Pi$ qui est le complémentaire de $T\theta$. Enfin, $T\Pi Z$, écart en longitude entre la tache et la Terre représentée par Z, s'obtient dans le triangle rectilatère $T\Pi Z$ ($\Pi Z = 90°$) par : $\cos T\Pi Z = \frac{\sin ZT}{\sin T\Pi}$.

Remarque : l'angle $T\Pi Z$ a même mesure que l'arc $\theta Z$. Toutes ces méthodes de trigonométrie sphérique découlent des problèmes traités l'*Astronomie de marins* (Avignon, 1766) du P. Pezenas et auxquels Jérôme Lalande se réfère en 1775.

[19] Par : $\cos TT' = \cos T\Pi.\cos T'\Pi + \sin T\Pi.\sin T'\Pi.\cos T\Pi T'$

différents, trouver le tems de sa révolution ?

On trouvera par le Problème précédent les trois arcs de grand cercle TT', T'T'' et TT'', et prenant les cordes de ces trois arcs, on en formera un triangle rectiligne TT'T'' (fig. 7.) que l'on inscrira dans un cercle. Ce cercle sera un parallèle à l'Équateur de la rotation ; et l'on fera cette proportion : comme l'angle TcT'', double supplément de TT'T'', est à 360° : : ainsi le tems écoulé entre la première et la troisième observation , est au tems de la révolution de la tache. Le demi-diamètre CT de ce petit cercle sera le sinus de la distance de la tache au pôle de la rotation[20].

/264v/

[Fig. 8.]

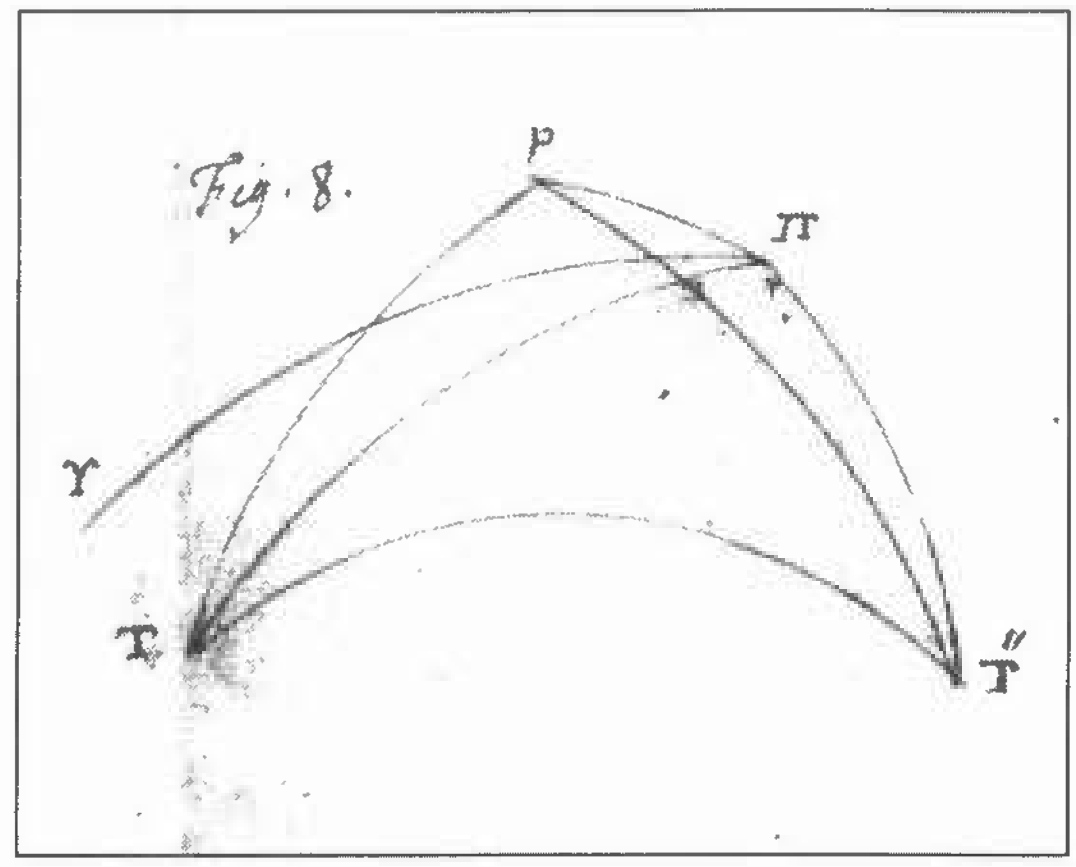

Problème 6.

Les mêmes choses étant supposées, trouver la longitude et la latitude du pôle de rotation, et l'inclinaison de son axe ?

[20] Entraînée par la rotation du Soleil, la tache T se déplace sur un parallèle pour cette rotation. Ce parallèle est un petit cercle σ sur Σ sur lequel se trouvent les trois positions T, T', T'' observées de la tache. Mais, par ce qui précède, on connaît les mesures des arcs TT', T'T'' et TT'', non comme arcs de σ mais comme arcs de grand cercle de Σ. Si l'on désigne par a, b,c ces trois mesures d'arcs, les cordes TT', T'T'' et TT'' seront proportionnelles à sin(a/2), sin(b/2), sin(c/2), ce qui permet de tracer un triangle TT'T'' semblable à ce triangle réel, de sorte que c'est par une méthode purement graphique qu'est finalement déterminée la rotation sidérale du Soleil.

Soient du pôle p de rotation (fig. 8.) les arcs pT, pT", pΠ, aux positions T et T" et au pôle Π de l'Écliptique. On a dans le triangle isocèle TpT", les trois cotés, aussi bien que dans le triangle TΠT". On aura donc les angles T"Tp et T"TΠ , et par conséquent les angles ΠTp, ΠT"p ; et dans les triangles pTΠ , pT"Π , ayant les deux cotés et l'angle compris, on aura la base pΠ , et les angles pΠT, pΠT". On aura donc l'inclinaison pΠ de l'axe de rotation à l'axe de l'Écliptique, et les différences en longitude entre le pôle p et les positions T, T" de la tache. Si la tache est en latitude sud, on prendra p pour le pôle Sud.

Exemple 1.

Le P. Poczobut jésuite polonais a fait en novembre 1762 les observations suivantes dans l'observatoire de Marseille. Le 15. $9^{bre}$ 1762 à 10 heures du matin, il découvrit une grande tache qui revint ensuite en décembre. Il l'observa le matin à $10^{h}\frac{1}{4}$, en faisant passer les deux bords du Soleil et la tache par le fil vertical de la lunette du quart de cercle, et mesurant la distance de cette tache au bord inférieur du Soleil par le moyen du micromètre. Il trouva le diamètre vertical du Soleil = 1280 parties du micromètre, et la distance de la tache au bord inférieur = 428 parties. La valeur de chaque partie a été trouvée = 1".31"',149. Le centre de la tache passa au fil vertical 2'.2" après le premier bord. Il l'observa de même le soir à $2^{h}\frac{3}{4}$. Nous nous bornerons aux observations qu'il a faites à midi les 15, 16, 18, 20, 21 et 22. $9^{bre}$, et nous choisirons celles qui nous paroîtront les plus exactes[21].

15. $9^{bre}$ 1762 à midi

Bord occidental du ☉ au fil méridional.........................12h. 3'. 59" de

[21] Données intéressantes au-delà du problème scientifique : remarquons l'activité régulière et obstinée des jésuites marseillais en ce mois de novembre 1762, alors que le Parlement de Paris a déjà condamné les jésuites et que celui d'Aix fait peser de graves menaces sur les conditions d'existence des jésuites provençaux.

l'horloge

La tache au même fil.....................................................12 . 5. 59

Bord oriental du Soleil au même fil...................................12 . 6. 18

Diamètre horizontal en tems................................................ 0 . 2. 19

Distance horizontale de la tache au premier bord en tems....0 . 2. 0.

en degrés du parallèle................................................30'.

Hauteur apparente du bord supérieur du Soleil, 28°. 24'. 45" ce qui donne la déclinaison du centre, 18°. 36'. 7".

Le rayon est au cosinus de la déclinaison, comme le sinus de 30', au sinus de la distance de la tache au bord occidental, 28'. 26". Ce qui donne la distance horizontale au centre, 12'. 11" = 731".

/265r/ [fig. 4. et fig. 5.]

La distance verticale a été trouvée sous le centre par le micromètre = 63" $\frac{1}{2}$.

La distance horizontale est à la verticale, comme le rayon est à la tangente de TCA (fig. 5.) = 4°. 57'. 33". Le sinus de cet angle est à la distance verticale, comme le rayon à la distance directe, 733" $\frac{3}{4}$ = 12'.13" $\frac{3}{4}$. Le demi-diamètre apparent est à la distance directe, comme le rayon est au sinus de CTK (fig. 4.) = 48°. 50'. 8". D'où retranchant cette distance directe, on trouve l'arc ZT = 48°.37'.54".

16. $9^{bre}$ 1762 à midi

Diamètre vertical du Soleil = 1286 parties du micromètre = 32'. 33". 36'''.

Bord occidental du ☉ au fil méridional.......12h. 7'. 38" de l'horloge

}differ. 1'.49". en degrés 0°.27'.15".

La tache au même fil.................................12 . 9. 27

Bord oriental [du Soleil] au même fil..........12 . 9. 55

Distance de la tache au bord inférieur par le micromètre = 643 parties.

Distance verticale au centre = 0.

Distance horizontale directe et réduite = 539".

Angle CTK = 33°.28'. 59". ZT = 33°.19'.14".

18. $9^{bre}$ 1762 à midi

Bord occidental du ☉ au fil méridional................12h. 6'. 9" de l'horloge

La tache au même fil..........................................12 . 7. 34

}differ. 1'.25".

Bord oriental [du Soleil] au même fil..................12 . 8. 26

Diamètre du ☉ en tems 2'. 17".

Distance horizontale réduite, 233" vers l'Orient.

Distance de la tache au bord inférieur 706. Diamètre vertical 1285.

Distance verticale au dessus du centre, 63 1/2 parties du micromètre = 106".

Donc la distance directe = 256". TCA = 24°.27'.45". CTK = 15°.22'. 3". ZT = 15°. 17'. 47".

20. $9^{bre}$ 1762 à midi

Bord occidental du ☉ au fil ......................12h. 7'. 14" de l'horloge

La tache au même fil...............................12 . 8. 11

}differ. 0'. 57".

Bord oriental [du Soleil] au même fil........12 . 9. 32

Distance horizontale en tems, 0'.12" vers l'Occident, en degrés 3'.0".

Demi-diamètre vertical 643 parties = 16'. 6". 48"'. Distance verticale de la tache 161 = 4'. 4". 33"' = 244" $\frac{1}{2}$.

Distance horizontale réduite 2'.49". = 169". Directe = 297".

TCB = 55°.17'. 34". ZT = 17°.36'.54".

21. $9^{bre}$ 1762 à midi

Bord occidental du ☉ au fil ........................12h. 7'. 36" de l'horloge

La tache au même fil..................................12 . 8. 19

}differ. 0'. 43".

Bord oriental [du Soleil] au même fil..........12 . 9. 54

Hauteur apparente du bord supérieur 26°.59'. 44". Déclinaison 20°.1'. 10".

Distance horizontale réduite 366". Verticale 294". Directe 469" $\frac{1}{2}$.

TCB = 38°.46'.33". ZT = 28°. 41'. 13".

/265v/

22. $9^{bre}$ 1762 à midi

Bord occidental du ☉ au fil ...............................12h. 8'. 5" de l'horloge

La tache au même fil........................................12 . 8. 36

}differ. 0'. 31".

Bord oriental [du Soleil] au même fil................12 .10. 23

Demi-diamètre en tems 1'. 9".

Hauteur apparente du bord supérieur 26°.46'. 26".37"'.

Déclinaison 20°.14'. 36".

Demi-diamètre vertical 643 parties = 16'.16".47"'. Distance horizontale 535".

Distance au bord inférieur 874 part. = 22'. 7". 44"'.

Distance verticale au centre = 5'. 50". 57"'. directe 639",8.

TCB = 33°.16'. 4". ZT = 40°.44'.50".

Le 15. $9^{bre}$ à midi, la tache étoit presque à l'extrémité du disque.

L'erreur d'une minute dans l'observation en produit une d'environ 7 à 8 degrés dans le calcul de TZ, [ratures] lorsque la tache est près du centre du disque, et une beaucoup plus grande lorsqu'elle est près des bords ; puisque dCT : dTZ : : comme 2 minutes : 15 degrés environ, ou comme 2" à 15'.

L'erreur dans les angles TCA ou TCB est la moindre, lorsque ces angles approchent plus de 45 degrés. Ainsi, dans le choix des trois observations, il faut préférer celles où ces angles s'éloignent le moins de 45°. Nous sommes donc forcés d'abandonner les deux premières observations, où cet angle ne surpasse pas 5 degrés, et de choisir les trois suivantes.

Ces sortes d'observations ne devraient pas se faire avec la lunette du quart de cercle, où l'on peut se tromper de 4 ou 5 secondes ; ce qui, comme on vient de le voir, doit produire une erreur d'un demi-degré (insertion : et quelquefois d'un degré entier) dans la route d'une tache du Soleil.

Il faudrait y employer une lunette de 15 à 20 pieds, et même le micromètre objectif de 40 pieds de foyer. Car si l'on place ce micromètre horizontalement à midi, par le moyen d'un petit niveau, on aura fort exactement la distance de la tache au bord oriental ou au bord occidental du Soleil ; et si on les place tout de suite verticalement, on aura la distance au bord supérieur ou au bord inférieur, et l'on ne sera exposé tout au plus qu'à une erreur d'une seconde.

[fig. 9.]

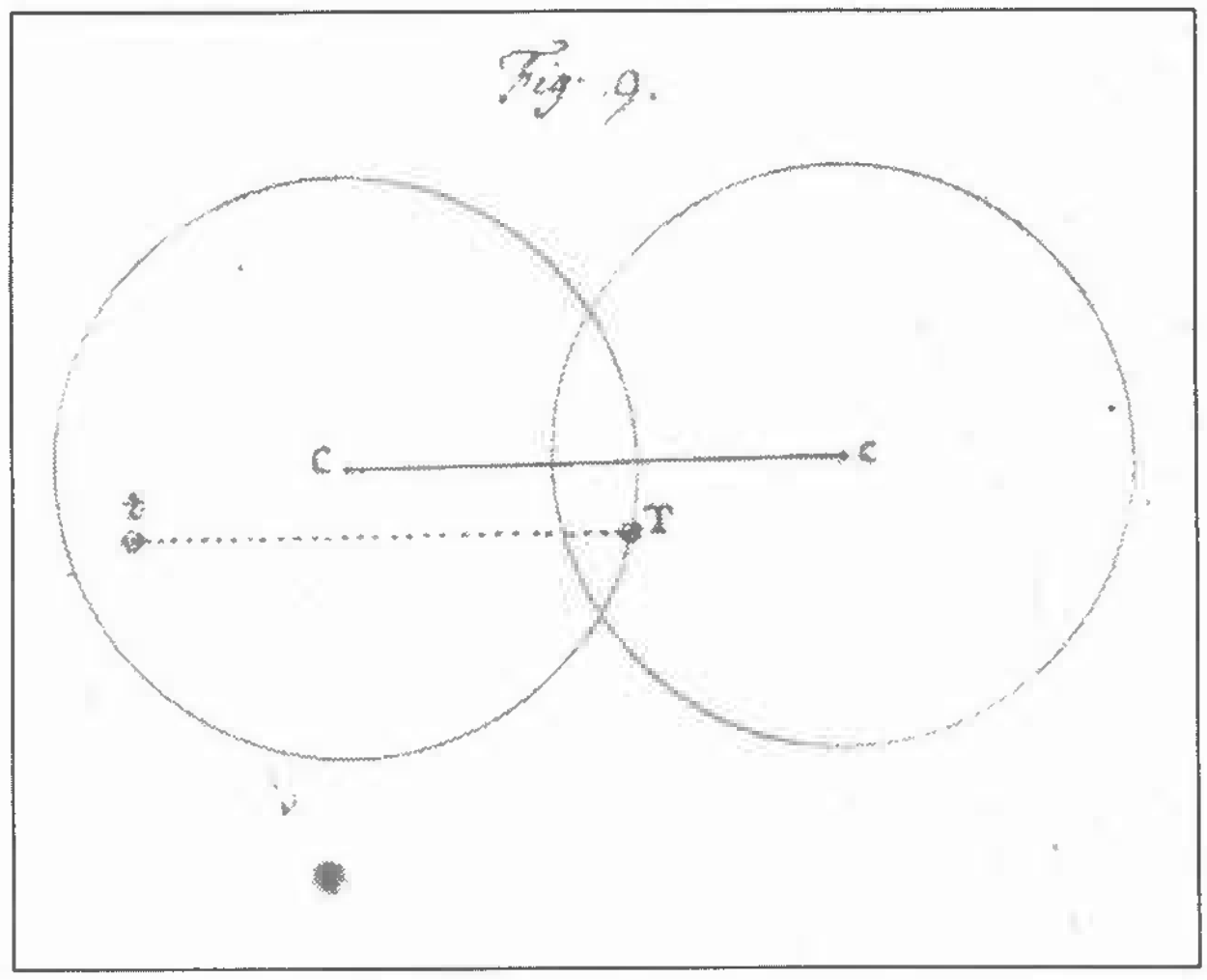

La ligne Cc qui joint les centres des deux images du Soleil (fig. 9.) étant horizontale, et le micromètre se mouvant horizontalement, dès que la tache T sera coupée par l'un des bords de l'autre image du Soleil, le micromètre donnera la distance tT de la tache à ce bord, puisque cette distance est toujours égale au chemin Cc qu'ont fait les centres des deux images en s'éloignant et que donne le micromètre. Il en sera de même pour la distance verticale de la même tache au bord supérieur ou inférieur.

[fig. 6.]

Voici le calcul des observations du 18, 20, et 22. $9^{bre}$ qui nous paroissent les plus exactes.

Celle du 18 donne TP = 80°.36'.14", et TPZ = 12°.7'.30" $\frac{1}{2}$. Celle du 20, T'P = 78°.16'. 17". et T'PZ' = 13°. 14'. 11". Celle du 22. donne ZZ" = 3°.56'.33".

AZ" = 59°.35'.16". T"P = 76°. 33'. 13". T"PZ" = 38°. 50'. 8".

/266r/

[fig. 7.]

Le triangle TΠT' donne la corde TT' = [0],46621056. T'ΠT" donne la corde T'T" = [0],46601918, et TΠT" donne TT" = [0],90591843.

Ces trois cordes donnent l'angle T' = 152°. 42'. 36" et TCT" = 54°. 34'. 48". Or cet angle est à 360°. comme 4 jours sont **au tems de la révolution, 26 jours, $9^h$ 11'. 31"**[22].

Les triangles BCT', DCT' donnent le sinus CT' de la distance au pôle, ou l'arc Tp (fig. 8.) = 81°. 5'. 17". Les triangles sphérique TpT", TPT" donnent T"Tp ou TT"p = 85°.25'.49".6. T"TΠ = 80°. 9'. 7",3. pTΠ = 5°. 16'. 42",". **Le triangle pTΠ donne pΠ = 5°. 14'.0"** [23]. TΠp = 94°. 53'. 55",6. D'où résulte ♈Πp *in antecedentia* 51°. 21'. 43". Ce qui étant soustrait de 12 signes, donne la longitude $10^s$. 8°. 38'. 17". [24]

[22] Souligné par nous ; résultat des observations des jésuites marseillais en 1762.

[23] *Ibid.*

[24] À cette époque, les astronomes comptent encore les longitudes écliptiques à partir du Point Vernal en divisant l'écliptique de 360° en douze signes d'étendue 360°/12 = 30°. Il faut donc lire dans la suite : $10^s$.8°.38'.17". = 308°.38'.17" à partir du point vernal (ici dans le sens rétrograde, *in antecendentia*).

Exemple 2°.

Dans *l'Histoire Céleste* de Mr. Le Monnier p. 206, on trouve une observation curieuse de Mr. Picard en 1676 sur une tache du Soleil qu'il apperçut (sic) le 26 Juin[25]. Il l'observa aussi le 28 Juin et 2 Juillet[26] ; mais il y a dans cette histoire une faute d'impression qui est aisée à corriger. L'observation marquée du 1er Juillet ne peut être que du 2e. car la différence des hauteurs du Soleil du 30 Juin au 1er Juillet y est marquée 8'.43", et elle ne peut être que de 3 ou 4 minutes. La hauteur le 30 fut 64°. 35'. 55" ; et le 1er ou 2e Juillet 64°. 27'. 12". La différence est 8'.43"., qui ne convient qu'à l'intervalle de deux jours[27]. Il fit donc les observations suivantes :

26 Juin 1676 à midi

1er bord du ☉ au méridien.. $0^h$ 2' 51"
la tache .............0. 4. 41 $\frac{1}{2}$
2e bord...............0. 5. 16
demi-diamètre en tems.........0. 1. 12 $\frac{1}{2}$

hauteur du bord Super.☉.... 64°.48'. 25"
de la tache .....64. 29. 0
distance au bord Sup............0. 19. 25
demi-diamètre du ☉ ............0. 15. 47
~~orient~~ distance au centre........0. 3. 38 = 218"
distance horizontale au 2e bord, 0'. 34" $\frac{1}{2}$
au centre 0. 38., en degrés 9'.30"

27 Juin à midi
1er bord du ☉ au méridien ........$0^h$ 3' 20"
la tache................0. 4. 47

[25] P.C. Le Monnier, 1741, *op. cit.*, p. 206 suivie d'une planche des dessins de la tache entre le 26 juin et le 4 juillet 1676.

[26] Il s'agit de l'observation indiquée pour le 1er juillet que Pezenas corrige dans la suite comme devant être datée du 2 juillet.

[27] Ces observations se trouvent chez Le Monnier, 1741, *op. cit.,* p. 207.

2e bord....................0. 5. $37\frac{1}{2}$

demi-diamètre en tems.............0. 1. $8\frac{3}{4}$

distance horizontale au 2e bord, 0'. 50" $\frac{1}{2}$

au centre 0. 18. $\frac{1}{4}$

hauteur du bord Super.$^{\odot}$........ 64°.46'. 0"
de la tache .....64. 26.15.
distance au bord Sup.................0. 19.45
demi-diamètre du $\odot$ .................0. 15.47
distance au centre......................0. 3.58

/266v/

28 Juin

1er bord $\odot$ au méridien ..........$0^h$ 3' 38"
la tache ............0. 4. 48
2e bord................0. 5. $55\frac{1}{2}$

demi-diamètre ............0. 1. 8"$\frac{3}{4}$

distance horizontale au 2e bord, 1'. 7" $\frac{1}{2}$

au centre 0. 1 $\frac{1}{4}$
en degrés = 19". vers l'orient

hauteur du bord Super.$^{\odot}$........ 64°.43'. 0"
de la tache .....64. 23. 0
distance au bord Sup.................0. 20. 0.
demi-diamètre du $\odot$ .................0. 15. 47
distance au centre......................0. 4. 13 = 253"

2. Juillet

Le 1er bord a passé au méridien 0'. 15" avant la tache
Demi-diamètre.............................1. 8 $\frac{3}{4}$

distance horizontale à l'occident....0. 53 $\frac{3}{4}$, en degrés 13'. 26" $\frac{1}{4}$
hauteur du bord supérieur............64°. 27'. 12"
tache..............64. 7. 10
distance..................0. 20. 2
Demi-diamètre du ☉ .....................0. 15. 47
Distance au Sud.............................0. 4. 15.

Choisissons les observations du 26 et 28 Juin, et du 2 juillet. La longitude du Soleil étoit le 26 Juin à midi $3^s$. 5°. 27'. 15". et par conséquent celle de la Terre $9^s$. 5°. 27'. 15", et sa distance à la Balance ou ♎Z = 95°. 27'. 15". ZZ' pour 2 jours = 1°. 57'. 25". Donc ♎Z' = 97°. 24'. 40". Z'Z" pour 4 jours = 3°. 54'. 50". Donc ♎Z" = 101°. 19'. 30". Les distances au pôle nord ZP = 113°. 22'. 0". Z'P = 113°. 16'. 17". Z"P = 112°. 58'. 43".

26 Juin 1676

Le rayon est au cosinus de la déclinaison, comme la distance horizontale 570" à la distance réduite ; et cette distance est à la verticale 218", comme le rayon est à la tangente de TCA = 22°.37'.2". Le sinus de cet angle : à la verticale : : R : à la distance directe 566" = 9'. 26". Le demi-diamètre 15'. 47" est au rayon, comme cette distance est au sinus de l'angle CTK = 36°.46'.7". Ce qui donne ZT = 36°.36'.41", par la soustraction de la distance directe 9'. 26".

28 Juin 1676

Le rayon est au cosinus de la déclinaison, comme la distance horizontale vers l'Orient 19", est à la distance réduite. Celle ci est à la verticale 253", comme le rayon à la tangente de T'CA = 86°.3'.12". Le sinus de cet angle : à la verticale : : R : à la directe 4'. 14". Le demi-diamètre : à cette distance : : R : au sinus de CT'K = 15°.31'. 59". D'où soustrayant 4'.14", on a Z'T' = 15°.27'.45".

/267r/

2. Juillet 1676

La même méthode donne T"CB à l'occident = 18°. 57'. 34". CT"K = 55°. 58'. 27". La distance directe CT" = 13'. 5", et Z"T" = 55°. 45'. 22".

26 Juin

Le triangle ΩPZ donne l'angle ΩZP = 92°.21'.51",4. La tache T étant sous la ligne horizontale, on a TZP = 90° + TCA = 112°. 37'. 2". D'où retranchant ΩZP, on aura ΩZT = 20°.15'.10",6. TZ = 36°. 36'. 41".

TΠ = 101°.54'. 51",3. et TΠZ = 34°. 52'. 40",7.

28 Juin.

ΩZ'P = 93°. 12'. 25". La tache étant toujours sous l'horizontale, on aura T'Z'P = 90° + T'CA = 176°.3'.12". D'où retranchant ΩZ'P, on aura T'Z'θ =82°.50'.47''. T'Π = 105°.20'.17".

R : cos ΩZ'T' : : Tang T'Z' : Tang Z'θ = T'ΠZ' = 1°.58'.22" $\frac{1}{2}$.

2. Juillet

On trouvera de même ΩZ"P = 94°. 52'. 18", et T"Z" étant sous l'horizontale à l'occident de Z", on a T"Z"P = 90° +T"CB = 108°.57'.34". D'où je retranche le supplément de ΩZ"P, ou l'angle ♈Z"P = 85°. 7'.42"., pour avoir T"Z"θ = 23°.49'.52''. R : sin T"Z" : sin T"θ" latitude, qui donne ΠT"=109°.30'.42".

R : cos ♈Z"T" : : tang T"Z" : tang Z"θ = T"ΠZ" = 53°.20'.40",6.

Nous avons donc par les calculs précédents ΩΠZ = 95°.27'.15"

- TΠZ = 34. 52. 40,7

ΩΠT = 60. 34. 34,3

---

ΩΠZ' = 97. 24. 40.

et T' étant à l'orient de Z', il faut soustraire T'ΠZ' = 1. 58. 22,5

ΩΠT" = 154. 40. 10,6.

Otez ΩΠT' = 95. 26. 17,5

T"ΠT' = 59. 13. 53,1

---

ΩΠT' = 95. 26. 17,5

- ΩΠT = 60. 34. 34,3

TΠT' = 34. 51. 43,2

T'ΠT" = 59. 13. 53,1

TΠT" = 94. 5. 36,3

/267v/

Le triangle TΠT' donne TT' = 34°. 0'. 57"

T'ΠT" ... T'T" = 56. 24. 6.

TΠT" .... TT" = 89. 49. 17.

Le triangle rectiligne sous ces trois cordes donne TT'T" = 133°. 17'. 27",6

TCT" = 93. 25. 4,8 = 93,418

Par conséquent la révolution, de **23 jours 3 heures**[28].

Le sinus $\frac{1}{2}$ TCT" est au rayon, comme $\frac{1}{2}$ TT" au sin. de Tp = 75°. 55'. 1",3.

Le triangle sphérique TpT" donne pT"T = 75°. 43'. 20".

Le triangle TΠT" donne TT"Π = 77. 24. 47,2

donc pT"Π = 1. 41. 27,2

Le triangle pT"Π donne pΠT" = 163. 8. 35.

- ΩΠT" = 154. 40. 11.

---

[28] Souligné par nous : résultat provenant donc de l'application de la méthode de Pezenas aux observations corrigées de Picard publiés par Le Monnier.

---

$\Omega\Pi T$ = 8. 28. 24.

Ce qui ôté de 360°, on a la longitude du pôle p = $11^s$. 21°.31'. 36".

Le même triangle donne l'inclinaison pΠ = 5°. 39'. 42" ½

[fin du mss ADH, D.128]

# ANNEXE 1

Reproduction du mémoire « Nouvelle théorie des taches du Soleil » du P. Pezenas imprimé dans les *Mémoires de mathématique et de physique, présentés par des savants étrangers & lus en séance*, dits « recueils des savants étrangers », tome VI, 1776, pp. 318-322 (+ pl.)

# *NOUVELLE THÉORIE DES TACHES DU SOLEIL.*

Par le P. PÉZENAS, Hiſtoriographe du Roi, à Marſeille.

## *DÉFINITIONS.*

J'APPELLE *Zénith* du Soleil le point $Z$ *(fig. 1)* le plus proche du centre de la Terre ou de l'œil $O$ de l'Obſervateur : ce point eſt toujours dans la ligne qui joint les deux centres $C$ & $O$, & cette ligne eſt toute entière dans l'Écliptique; $AZC$ eſt le demi-cercle du globe du Soleil, $CA$ ſon demi-diamètre réel, perpendiculaire au rayon viſuel $OC$; $OD$ eſt un autre rayon viſuel, qui touche le quart-de-cercle $ADZ$ en $D$ & qui eſt par conſéquent perpendiculaire au demi-diamètre $CD$; l'arc $DA$, qui meſure l'angle $DCA$, complément de $ZCD$, eſt auſſi la meſure du demi-diamètre apparent $COD$, complément de $ZCD$.

Si l'on imagine un petit cercle dont le diamètre ſoit $DG$ parallèle à $AB$ & qui ſoit perpendiculaire au plan $BZA$; ce petit cercle repréſentera le diſque du Soleil; ſon centre $F$ ſera la projection du zénith $Z$; les lignes droites $DF$, $FG$ ſeront les projections des arcs $DZ$, $GZ$, & par la même raiſon tous les demi-diamètres du diſque du Soleil, ſeront les projections de tous les arcs qui ſe coupent en $Z$.

## PROBLEME I.

*Trouver la grandeur de l'arc* ZG *compris entre le zénith* Z *& une tache quelconque* T, *dont on a la diſtance apparente* COT, *& que l'on ſuppoſe placée ſur la ſurface du Soleil.*

Prolongez $OT$ juſqu'au diamètre réel en $k$; vous aurez ces proportions; le ſinus de $COT$ eſt au ſinus de $CTO$ ou de $CTK$ :: $CTZ$ ou $CD$ : $CO$ :: (à cauſe des triangles rectangles $CDO$ & $CDF$ qui ont un angle commun $FCD$) $CF$ : $CD$

comme le ſinus de l'arc $DA$ eſt au rayon; donc le ſinus du demi-diamètre apparent ou de l'arc $DA$ eſt au rayon, comme le ſinus de la diſtance apparente de la tache ou zénith, eſt au ſinus de l'angle $CTK$, extérieur au triangle $CTO$; donc en retranchant de cet angle $CTK$ l'angle intérieur $COT$, on aura l'angle $ZCT$ ou l'arc $ZT$; l'obſervation donne le demi-diamètre apparent $COD$ & la diſtance apparente $COT$ de la tache $T$ au zénith, ſoit qu'elle ſe trouve dans l'arc $ZTD$ ou dans un autre grand cercle du globe du Soleil; on aura donc l'angle $CTK$, qui eſt toujours plus grand que $ZCT$; ſi l'œil étoit à une diſtance infinie du Soleil, la projection $Ft$ de $ZT$ ſeroit égale au ſinus de cet arc; mais on ne peut pas admettre cette hypothèſe, & $Ft$ eſt toujours plus grand que le ſinus de $ZT$ ſous le rayon $CT$.

## COROLLAIRE.

Cette méthode s'applique naturellement aux Satellites d'une Planète, en diſant, comme le ſinus de la plus grande digreſſion du Satellite eſt au rayon; ainſi le ſinus de ſa diſtance apparente au centre, eſt au ſinus de l'angle $CTK$, ce qui donne l'arc $ZT$.

## PROBLEME II.

*Trouver la valeur de l'angle formé au zénith par le grand cercle* ZD, *& par l'arc* ZT.

L'angle rectiligne $TCA$ *(fig. 2)* formé au centre du diſque par la diſtance apparente $TC$ de la tache à ce centre & par le demi-diamètre horizontal $CA$, eſt égal à l'angle ſphérique formé par l'arc du grand cercle $ZT$, dont la projection eſt $CK$ *(fig. 1)* & par l'arc $ZD$ dont la projection eſt $CA$; car ces deux angles ſont meſurés par le même arc de cercle de projection dont $C$ eſt le centre & dont le pôle eſt le zénith; or les obſervations donnent les lignes $TC$ & $CA$ *(fig. 2)* & l'angle $TCA$; donc on aura par obſervation l'angle ſphérique requis.

## PROBLEME III.

*Trouver la latitude & la longitude d'une tache du Soleil.*

Soit $V$ *(fig. 3)* le plus prochain équinoxe dans le globe du ſoleil, $VZ$ l'écliptique & la diſtance de la Terre à cet équinoxe; $P$ le pôle de l'équateur dans le méridien $PCZ$; on a les trois côtés du triangle $VZP$, ſavoir, $VP$ de 90 degrés, la diſtance $ZP$ du lieu de la Terre au pôle nord & la diſtance $VZ$ au plus prochain équinoxe; on aura donc l'angle $VZP$; d'où retranchant l'angle $TZP$, dont on a trouvé le complément $TZD$ (*fig. 1*) par le ſecond Problème, où ajoutant $TZP$ ſi la tache eſt plus occidentale que $ZP$, on aura l'angle $VZT$ ou $VZT'$.

Soit du pôle $\pi$ de l'écliptique le cercle de latitude $\pi T\theta$ perpendiculaire à l'écliptique en $\theta$; dans le triangle rectangle $ZT\theta$, on connoît $TZ$ & l'angle $TZ\theta$; on aura donc la latitude $T\theta$ & ſa diſtance $T\pi$ au pôle de l'écliptique; & dans le triangle $TZ\pi$, connoiſſant les trois côtés $TZ$, $T\pi$ & $Z\pi$ de 90 degrés, on aura l'angle $T\pi Z$, qui eſt la différence en longitude entre la tache $T$ & la Terre $Z$.

## PROBLEME IV.

*Les latitudes & les longitudes d'une même tache étant données; en deux temps différens, trouver l'arc de grand cercle, qui joint ces deux poſitions.*

Le zénith $Z$ ſe mouvant avec la Terre, & $Z$ étant le lieu de la Terre dans la première obſervation, $Z'$ ſera le lieu de la Terre dans la ſeconde obſervation, ſi l'arc $ZZ$ eſt égal à l'accroiſſement de la longitude de la Terre entre les deux obſervations; on trouvera donc, comme dans le troiſième Problème, l'arc $\pi T'$, & l'angle $Z'\pi T'$, qui étant combiné avec les angles $T\pi Z$, $Z\pi Z'$, donnera l'angle $T\pi T'$; donc connoiſſant les deux côtés $T\pi$, $T'\pi$ & l'angle compris $T\pi T'$, on aura la baſe $TT'$ du triangle $T\pi T'$.

## PROBLEME V.

*Les latitudes & les longitudes d'une même tache, étant données en trois temps différens, trouver le temps de ſa révolution & la diſtance à ſon pôle de rotation.*

On trouvera par le Problème précédent les trois arcs de grand cercle $TT'$, $T'T''$, $TT''$, & prenant les cordes de ces trois arcs, on en formera un triangle rectiligne $TT'T''$ (*fig. 4*) que que l'on inſcrira dans un cercle; ce cercle ſera un parallèle à l'équateur de la rotation, & l'on ſera cette proportion, comme l'angle $TCT''$ double ſupplément de $TT'T''$ eſt à 360 degrés comme le temps écoulé entre la première & la troiſième obſervation eſt au temps de la révolution de la tache.

Le demi-diamètre $CT$ de ce petit cercle, ſera le ſinus de la diſtance de la tache au pôle de ſa rotation.

## PROBLEME VI.

*Les mêmes choſes étant ſuppoſées, trouver la longitude & la latitude du pôle de rotation & l'inclinaiſon de ſon axe.*

Soient du pôle $p$ de rotation *(fig. 5)* les arcs ou diſtances $pT, pT'', p\pi$ aux poſitions $T$ & $T''$ & au pôle $\pi$ de l'écliptique; on a dans le triangle iſoſcèle $TpT''$, les trois côtés, auſſi-bien que dans le triangle $T\pi T''$; on aura donc les angles $T''Tp$ & $T''T\pi$, & par conſéquent les angles $\pi Tp$, $\pi T''p$; & dans les triangles $pT\pi$, $pT''\pi$, ayant les deux côtés & l'angle compris, on aura la baſe $p\pi$ & les angles $p\pi T$, $p\pi T''$: on aura donc l'inclinaiſon $p\pi$ de l'axe de rotation & les différences en longitude entre le pôle $p$ & les poſitions $TT''$ de la tache; ſi ſa latitude eſt ſud, on prendra $p$ pour ſon pôle ſud.

### *REMARQUES.*

(1.) L'erreur d'une minute dans l'obſervation, en produit une d'environ 15 degrés dans le calcul de l'arc $TZ$; d'où il ſuit que ces ſortes d'obſervations ne devroient pas ſe faire avec la lunette du quart-de-cercle où l'on peut ſe tromper de 4 ou 5 ſecondes, ce qui doit produire une erreur d'un degré dans la route d'une tache du Soleil; il faudroit y employer une lunette de 15 à 20 pieds, & même le micromètre objectif de 40 pieds de foyer; car ſi l'on place ce micromètre horizontalement à midi, par le moyen d'un petit niveau, on aura fort exactement la diſtance de la tache au bord oriental ou au bord occidental du Soleil; & ſi on le place tout de ſuite verticalement, on aura ſa diſtance au bord ſupérieur ou au bord inférieur, & l'on ne ſera expoſé tout au plus qu'à une erreur d'une ſeconde.

(2.) Dans l'Hiſtoire Céleſte de M. le Monnier, *page 206*, on trouve une obſervation curieuſe de M. Picard, en 1676, ſur une tache du Soleil qu'il aperçut le 26 Juin; il l'obſerva le 28 Juin & le 2 Juillet, mais il y a dans cette Hiſtoire une faute d'impreſſion qu'on peut corriger aiſément; l'obſervation marquée du $1.^{er}$ Juillet, ne peut être que du 2, car la différence des hauteurs du Soleil du 30 Juin au $1.^{er}$ Juillet eſt marquée $8'\,43''$, & elle ne peut être que de 3 ou 4 minutes. M. Picard n'employa pas le micromètre, & en choiſiſſant les obſervations des 26 & 28 Juin & du 2 Juillet, je n'ai trouvé la révolution que de 23 jours 3 heures, l'inclinaiſon $5^{d}\,39'\,42''$ & la longitude du pôle $11^{ſ}\,21^{d}\,31'\,36''$.

(3.) Le P. Poczobut, Jéſuite Polonois, fit en Novembre 1762, dans l'Obſervatoire de Marſeille, pluſieurs obſervations ſur une grande tache qui revint en Décembre; il y employa le quart-de-cercle de la Marine, armé d'un bon micromètre; j'ai choiſi les obſervations des 18, 20 & 21 Novembre, qui m'ont paru les plus exactes; elles m'ont donné la révolution de 26 jours 9 heures, l'inclinaiſon $5^{d}\,14'$, & la longitude du pôle $10^{ſ}\,8^{d}\,38'$; cette tache qui avoit paru le 15 Novembre, 2 ſecondes de temps après le paſſage du $1.^{er}$ bord du Soleil, parut de nouveau le 11 Décembre, $2'\,16''$ après le même bord; c'eſt-à-dire 27 jours après, ce qui donneroit environ 25 jours à ſa révolution, ſelon la méthode ordinaire; cette méthode ſuppoſe que la révolution ſe fait autour de l'axe de l'écliptique, mais l'axe de rotation étant incliné de 5 à 6 degrés à l'axe de l'écliptique, la méthode ordinaire doit donner une révolution trop grande ou trop petite.

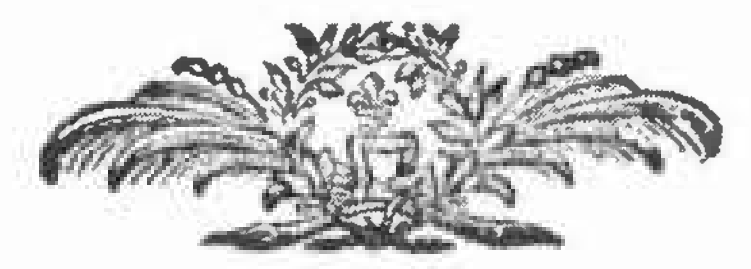

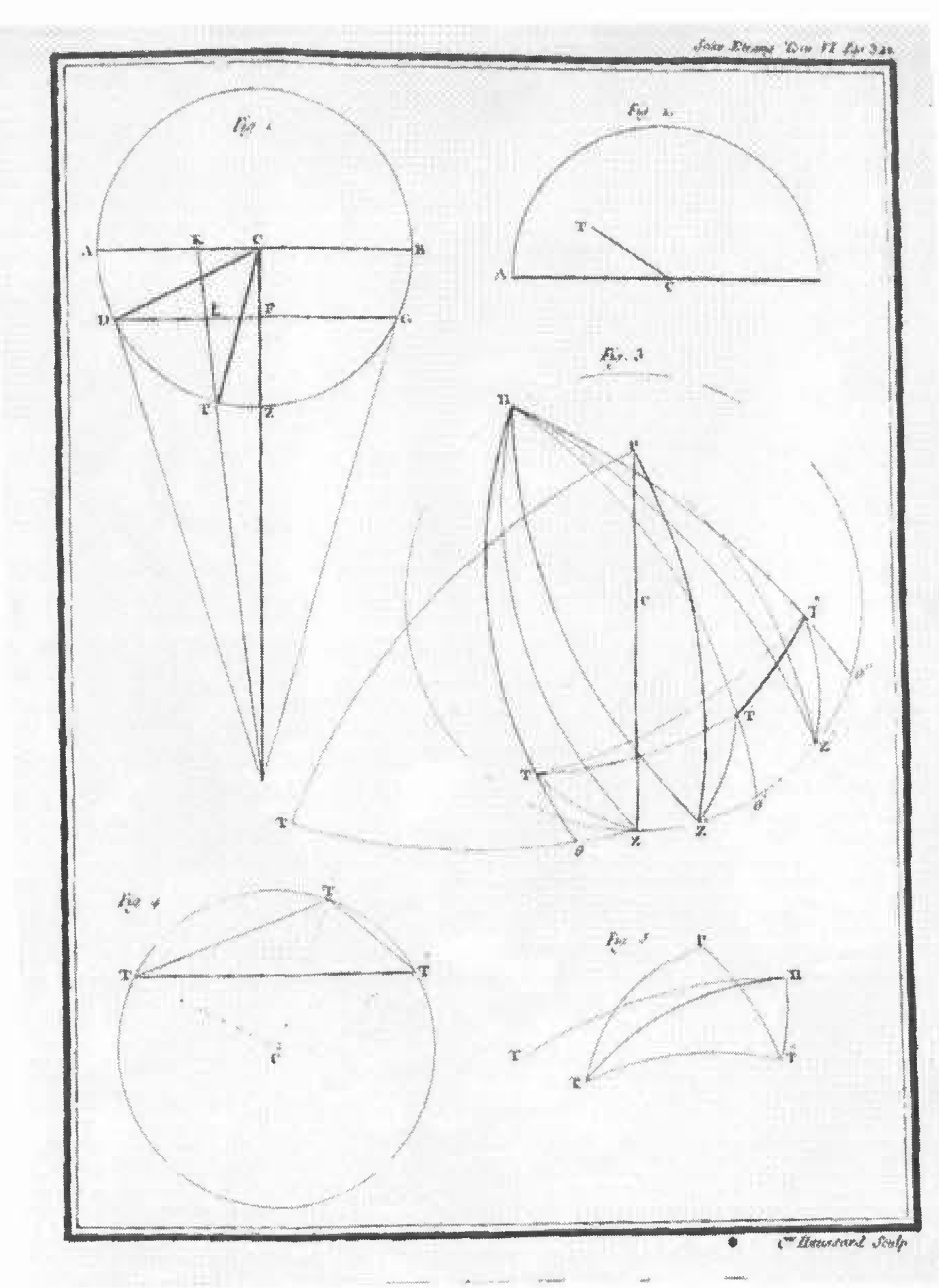

# ANNEXE 2

## Rapport de Joseph-Nicolas Delisle et du P. Alexandre-Guy Pingré sur le mémoire « Nouvelle théorie des taches du Soleil » présenté par le P. Pezenas, Procès-verbaux des séances de l'Académie royale des sciences, 20 août 1766

« MM. De Lisle et Pingré ont fait le rapport suivant de la nouvelle théorie des taches du Soleil par le père Pezenas.

L'auteur, après avoir dit qu'il donne le nom de zénith du Soleil au point de sa surface la plus proche de la Terre, à celui où la ligne qui joint les centres du Soleil et de la Terre rencontre la surface du Soleil, donne dans un premier problème le moyen de trouver sur la surface du Soleil la valeur de l'arc compris entre le zénith du Soleil & une tache quelconque. Dans un second problème, il s'agit de déterminer l'angle formé par l'arc susdit et un grand cercle passant par le zénith parallèlement à l'équateur terrestre. On en conclut dans le troisième problème la longitude et la latitude de la tache observée.

Le quatrième problème enseigne à trouver l'arc de grand cercle compris entre deux positions observées d'une même tache.

Dans le cinquième problème, on tire une méthode pour déterminer le temps de la révolution de la tache et sa distance à son pôle de rotation.

Enfin, dans le sixième problème, on détermine par les mêmes données la longitude et la latitude du pôle de rotation et l'inclinaison de son axe.

Le père Pezenas, avertit qu'une minute d'erreur en occasionne une de quinze degrés dans le mouvement de la tache ; qu'en conséquence, on ne peut être trop précis ni employer de trop bons instruments dans les observations différentes de la tache.

On pourroit ajouter que l'erreur est d'autant plus à craindre que la tache seroit observée à une plus grande distance de l'axe du Soleil ou même de son zénith. Le père Pezenas remarque ensuite une faute d'impression qui s'est glissée dans une observation curieuse de Mr. Picard, faite en seize cent soixante seize les vingt six et vingt huit juin et non pas le premier mars, mais le deux juillet ayant pour objet une tache du Soleil et rapportée dans l'*Histoire céleste* de M. le Monnier. Des différentes positions d'une tache observée en novembre mil sept cent soixante deux par le p. Poczobut, jésuite polonais, le p. Pezenas conclut la révolution du Soleil de vingt six jours neuf heures, l'inclinaison de son axe à celui de l'écliptique de cinq degrés & quatorze minutes, et la longitude du pôle dix signes huit degrés trente huit minutes.

Ce mémoire nous a parû curieux & instructif. La méthode qui y est proposée n'est peut être pas aussi nouvelle que son auteur paroît le croire. Nous avons remarqué bien des traits de ressemblance entre cette méthode et une de celles que M. de Lisle a proposées à l'occasion d'une tache observée par lui, il y a plus de cinquante trois ans, et qui sont imprimées dans les *Mémoires pour servir à l'histoire et au progrès de l'astronomie, de la géographie et de la physique*, à St Pétersbourg, in 4° page 140 et suiv.

Cependant, comme la ressemblance n'est pas complète, et que d'ailleurs le mémoire du p. Pezenas nous paroît précis, méthodique et très utile, nous croyons qu'il mérite d'être imprimé dans le recueil des mémoires présentés à l'académie par des Savants étrangers, en ajoutant une note sur l'épithète de nouvelle que le p. Pezenas donne à sa méthode.

Fait à Paris, le vingt du mois d'août. »

# ANNEXE 3

## Rapport d'Alexis Clairaut et de Jérôme Lalande sur le mémoire « Problème » présenté par Guillaume de Saint-Jacques de Silvabelle, Procès-verbaux des séances de l'Académie royale des sciences, le 15 février 1764.

« MM. Clairaut et Lalande ont fait le rapport suivant du problème de M. de Silvabelle.

Ce mémoire qui contient la solution de ce problème par le moyen de trois observations d'une tache du Soleil, déterminer le cercle qui décrit la tache ; l'auteur est M. de Silvabelle, dont l'académie a déjà vû et approuvé plusieurs mémoires. Le problème dont il s'agit ici est important pour la connoissance de la rotation du Soleil ou de la Lune, car il s'applique également à tous deux ; il sert à connaître tout à la fois l'inclinaison de l'équateur ou de l'axe de la planète sur l'écliptique, le lieu où cet équateur rencontre l'écliptique, et la distance de la tache qui a été observée par rapport à l'équateur de sa planète.

M. Delisle en a donné la solution par des opérations graphiques, dans ses mémoires imprimés à Petersbourg ; M. Mayer dans les actes de la société cosmographique de Nuremberg imprimés en 1748 en a donné une solution par approximation à l'occasion de la libration de la Lune, et sa solution est forte élégante et fort commode. Le P. Boscovich croyant qu'une solution directe étoit plus convenable qu'une approximation, en chercha une qu'il communiqua à M. Delisle et qui est encore manuscrite. Celle de M. de Silvabelle est aussi géométrique et aussi simple que celle du Père Boscovich ; M. de Silvabelle ne suppose que trois longitudes et latitudes vûes de la Terre, il trouve une équation générale entre l'inclinaison du cercle que la tache décrit, le mouvement de la Terre dans l'intervalle des observations, la distance de la tache à la Terre, sa latitude observée et la différence des longitudes ; d'où il est aisé de tirer par la substitution des nombres donnés par trois observations, toutes les inconnûes de ce problème.

Les astronomes ont besoin, il est vrai, d'une solution qui se rapporte au centre du Soleil et qui soit sous une forme différente ; mais cela n'empêche pas que la solution dont il s'agit ici ne soit très bonne et très digne d'un Géomètre tel que M. de Silvabelle ; nous croyons donc que ces solutions méritent d'être imprimées dans le recueil des mémoires présentés à l'académie. »

# ANNEXE 4

« Problème », mémoire de Guillaume Saint-Jacques de Silvabelle, publié dans les « Savants étrangers », tome V, 1768, pp. 631-634.

# PROBLÈME

Par M. DE SAINT-JACQUES DE SILVABELLE.

*TROIS obſervations d'une tache du Soleil étant données, déterminer le parallèle du Soleil que décrit la tache, & le temps de ſa révolution.*

## SOLUTION.

Soit $C$ le centre du Soleil, $E^\circ$, $E$, $E'$, les lieux obſervés de la tache; $T^\circ$, $T$, $T'$, les lieux de la Terre dans les trois obſervations; $e^\circ$, $e$, $e'$ ſont les projections des lieux $E^\circ$, $E$, $E'$ de la tache ſur le plan de l'écliptique: les lignes $T^\circ e^\circ$, $Te$ $T'e'$ marquent les longitudes de la tache.

Nous regardons ici, pour ſimplifier les expreſſions, les points $T^\circ$, $T$, $T'$ de l'orbite terreſtre comme placés dans un arc de cercle, dont le Soleil eſt le centre.

$DN$ eſt la ligne d'interſection du plan de l'écliptique & du plan du cercle que décrit la tache; $D$ eſt le point où le rayon $T^\circ C$ rencontre cette ligne.

Par le point $T^\circ$, ayant mené la ligne $T^\circ N$ perpendiculaire à $CT^\circ$, on élèvera ſur le même point la perpendiculaire $T^\circ \pi$ au point de l'écliptique; $\pi$ eſt le point où cette perpendiculaire rencontre le plan du cercle que décrit la tache, lequel ſe trouve déterminé par la poſition des trois points $\pi$, $N$, $D$.

Du point $e$, ayant abaiſſé ſur la ligne $CT^\circ$, la perpendiculaire $eB$, qui rencontre en $K$ la ligne des nœuds, on abaiſſera du point $T$ ſur cette ligne $eB$ la perpendiculaire $TF$, qui ſera parallèle au rayon $CT^\circ$.

Enfin on mènera la ligne $CG$ perpendiculaire à $Te$, & l'on tirera les lignes $CT$, $Ce$, $CE$, $EK$, $Ee$, $\pi N$.

Nommant 1 le rayon $CT^\circ$ ou $CT$, que l'on prend pour le ſinus total, $FB$ ſera égal au ſinus de l'arc $TT^\circ$, que nous nommerons $B$, & ſon coſinus $b$; $CG$ ſera le ſinus de l'angle $CTe$, que nous nommerons $G$, & ſon coſinus $g$.

Soit encore nommé $F$ le ſinus de l'angle $FTe$, & $f$ ſon coſinus, $L$ la tangente de l'angle $eTE$, ou de la latitude de la tache lorſqu'elle ſe trouve en $E$.

On nommera les lignes $T^{\circ} N$, $n$; $T^{\circ}\pi$, $\pi$; $Te$, $Z$; $CE$, $r$; $CD$, $x$.

On aura $(T^{\circ} D) = 1 + x$, $(eE) = L \times z$, $(eF) = F \times z$, $(TF) = f \times z$, $(eB) = (FB) - (eF)$, $= B - F \times z$, $(CB) = b - fz$, $(Ce) = \sqrt{[(g - z)^2 + GG]} = \sqrt{(1 - 2gz + zz)}$; donc $(CE)$ ou $r = \sqrt{(1 - 2gz + zz + LLzz}$; d'où l'on tire $z = \frac{g \pm \sqrt{[(rr - 1) \times (1 + LL) + gg]}}{1 + LL}$, dans laquelle valeur de $z$ il n'entre que des grandeurs données, $r$ étant ici le rayon du Soleil qui eſt connu.

C'eſt pourquoi, afin d'abréger les expreſſions, nous ferons $(eB) = a$, $(CB) = \beta$, $eE = \gamma$, puiſque dans les valeurs de ces lignes trouvées ci-deſſus, il n'entre que des grandeurs connues, & $z$ qui devient connu.

Les triangles ſemblables rectangles $DT^{\circ}N$, $DBK$ donnent $(BK) = \frac{n}{1 + x} \times (BD) = \frac{n}{1 + x} \times (\beta + x)$; d'où l'on tire $(eK)$ ou $(BK) - (Be) = \frac{n}{1 + x} \times (\beta + x) - a$; & les triangles ſemblables rectangles $NT^{\circ}\pi$, $KeE$, donnent $eE$ ou $\gamma = \frac{(T^{\circ}\pi)}{(T^{\circ}N)} \times (eK) = \frac{\pi}{n} \times [\frac{n}{1 + x} \times (\beta + x) - a]$; d'où l'on tire

$$\pi = \frac{n\gamma}{\frac{n}{1 + x} \times (\beta + x) - a}.$$

COROLLAIRE I.

L'équation qu'on vient de trouver pour l'obſervation du point point $E$, & qui eſt $\pi = \dfrac{n\gamma}{\frac{n}{1+x} \times (\beta + x) - \alpha}$, convient également à toute autre obſervation, en mettant au lieu de $(EB)$, $(CB)$, $(eE)$, ou $\alpha$, $\beta$, $\gamma$, les lignes analogues qui conviennent à cette obſervation, les grandeurs $n$, $\pi$, $x$ demeurant invariables & les mêmes pour toutes les obſervations.

Ainſi pour l'obſervation faite en $T'$ du lieu $E'$ de la tache, on aura $\pi = \dfrac{n\gamma'}{\frac{n}{1+x} \times (\beta' + x) - \alpha'}$, & pour l'obſervation faite en $T^{\circ}$ du lieu $E^{\circ}$, $\pi = \dfrac{n\gamma^{\circ}}{\frac{n}{1+x} \times (\beta^{\circ} + x) - \alpha^{\circ}}$

## COROLLAIRE II.

Si l'on compare entr'elles les deux premières valeurs de $\pi$ du corollaire précédent, on aura $n = \dfrac{(\alpha'\gamma - \alpha\gamma') \times (1 + x)}{\gamma \times (\beta' + x) - \gamma' \times (\beta + x)}$ & ſi l'on compare entr'elles la première & la troiſième valeur de $\pi$ du même corollaire, on aura $n = \dfrac{(\alpha^{\circ}\gamma - \alpha\gamma^{\circ}) \times (1 + x)}{\gamma \times (\beta^{\circ} + x) - \gamma^{\circ} \times (\beta + x)}$ & comparant ces deux valeurs de $n$, on aura

$$x = \frac{(\alpha^{\circ}\gamma - \alpha\gamma^{\circ}) \times (\beta'\gamma - \beta\gamma') + (\alpha'\gamma' - \alpha\gamma') \times (\beta\gamma^{\circ} - \beta^{\circ}\gamma)}{(\alpha'\gamma - \alpha\gamma') \times (\gamma - \gamma^{\circ}) + (\alpha^{\circ}\gamma - \alpha\gamma^{\circ}) \times (\gamma' - \gamma)},$$

ce qui donne la valeur de $x$, & conſéquemment on aura, par les équations ci-deſſus, les valeurs de $n$ & de $\pi$, & tout ſera connu dans le problème.

## COROLLAIRE III.

Par le moyen du problème précédent & des deux premiers corollaires, il eſt aiſé de réſoudre tout ce qu'on peut ſouhaiter là-deſſus.

Par exemple, pour déterminer le rayon $cE$ du parallèle que décrit la tache, il n'y a qu'à abaiſſer du centre $C$ du Soleil la perpendiculaire $CH$ à la ligne des nœuds $DN$.

Alors dans le triangle rectangle $CHD$, ſemblable au triangle $DT^{\circ}N$, qui eſt tout connu, connoiſſant le côté $CD$, on aura la valeur de $CH$.

Et ſi du centre $c$ du parallèle que décrit la tache, on mène par le point $H$ la ligne $CH$, on aura le triangle rectangle $CcH$, dans lequel il ſera aiſé de connoître, par le problème, l'angle $CHc$, qui eſt l'angle d'inclinaiſon du plan du cercle de la tache avec le plan de l'écliptique; car ſi du point $T^{\circ}$ on mène une perpendiculaire $T^{\circ}P$ 1 à la ligne des nœuds $DN$, & qu'on prenne cette perpendiculaire pour rayon, la ligne $T^{\circ}\pi$ ſera la tangente de cet angle; de ſorte qu'on connoîtra dans ce triangle $CHc$ les trois angles & le côté $cH$, ce qui donnera le côté $Cc$, que l'on auroit également pu trouver encore plus ſimplement par cette proportion, $T^{\circ}P$ eſt à $T^{\circ}\pi$ comme $cH$ eſt à $Cc$.

Dans le triangle rectangle $CcE$, connoiſſant l'hypothénuſe $CE$, qui eſt le rayon du Soleil, & φ le côté $Cc$, qui eſt la diſtance des centres du Soleil & du cercle que décrit la tache, on connoîtra le côté $cE$, qui eſt le rayon de ce cercle.

Il ſera auſſi aiſé de connoître les lignes $ee'$, $EE'$, puiſque $(ee')^2$ eſt égal à la ſomme des quarrés des différences des lignes $e'B'$, $eB$ & $cB$, $cB'$; de ſorte que dans le triangle iſocèle $cEE'$, on connoîtra les trois côtés, & par conſequent l'angle $EcE'$; d'où il ſera facile de conclure le temps de la révolution de la tache par cette analogie, l'angle $EcE'$ eſt à 360 degrés, comme le temps écoulé entre les deux obſervations en $E$ & $E'$ eſt au temps de la révolution de la tache.

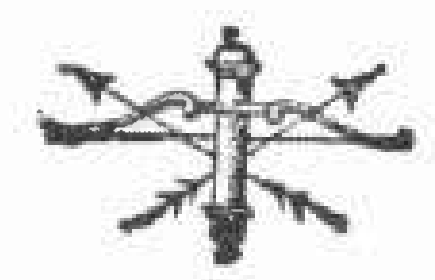